\documentclass[twocolumn,preprintnumbers,amsmath,amssymb,superscriptaddress,nofootinbib,longbibliography,prl]{revtex4-1}

\usepackage{graphicx}
\usepackage{bm}
\usepackage{dsfont}
\usepackage[usenames,dvipsnames]{xcolor}
\usepackage{pstricks}
\usepackage[tight]{subfigure}
\usepackage{verbatim}
\usepackage{units}
\usepackage{multirow}
\usepackage{enumitem}
\usepackage{mathrsfs}
\usepackage{leftidx}
\usepackage{xspace}
\usepackage{amsfonts,amssymb,amsmath}

\usepackage[a4paper]{hyperref}
\hypersetup{colorlinks=true,linktoc=all,linkcolor=blue,breaklinks=true,citecolor=blue,urlcolor=blue}
\usepackage[english]{babel}
\usepackage[normalem]{ulem}

\usepackage{hyperref}

\begin{document}
		
	
	\title{Readout of quantum screening effects using a time-dependent probe}
	
	\author{Nastaran Dashti}
	\affiliation{Department of Microtechnology and Nanoscience (MC2), Chalmers University of Technology, S-412 96 G\"oteborg, Sweden}

    \author{Matteo Acciai}
	\affiliation{Department of Microtechnology and Nanoscience (MC2), Chalmers University of Technology, S-412 96 G\"oteborg, Sweden}

	\author{Sara Kheradsoud}
	\affiliation{Physics Department and NanoLund, Lund University, S-221 00 Lund, Sweden}
	
	\author{Maciej Misiorny}
	\affiliation{Department of Microtechnology and Nanoscience (MC2), Chalmers University of Technology, S-412 96 G\"oteborg, Sweden}

	\author{Peter Samuelsson}
	\affiliation{Physics Department and NanoLund, Lund University, S-221 00 Lund, Sweden}
		
	\author{Janine Splettstoesser}
	\affiliation{Department of Microtechnology and Nanoscience (MC2), Chalmers University of Technology, S-412 96 G\"oteborg, Sweden}
	
	\date{\today}
	
	\begin{abstract}
In voltage- and temperature-biased  coherent conductors quantum screening effects occur if the conductor's transmission is energy-dependent. Here, we show that an additional ac-driven terminal can act as a probe for a direct readout of such effects, hitherto unexplored. We find that screening of charges induced by the static biases impacts already their standard \textit{linear} thermoelectric response coefficients due to \textit{nonlinear} effects when accounting for the frequency of the time-dependent driving. 
Those effects should be observable under realistic experimental conditions and can literally be switched on and off with the ac-driving. 
	\end{abstract}

	\maketitle

In recent years, there has been a growing interest in the field of 
nanoscale thermoelectrics~\cite{Benenti2017Jun}: by exploiting the features of nanoscale conductors -- such as their energy-dependent transmission properties, single-particle effects, and even quantum interference effects -- novel principles for electric heat-to-work conversion are currently explored. In contrast to analogous macroscopic devices, which are typically well characterized by their linear thermoelectric properties, the nonlinear response plays an important role for these nanoscale conductors, where applied temperature or voltage differences can easily be of the order of internal energy scales. However, the nonlinear operation of these devices goes along with complex quantum screening effects in the conductor, which impact their transmission properties~\cite{Christen1996Sep,Buttiker2005Jan,Buttiker1997b,Sanchez2004Sep,Sanchez2013Jan}.  Despite their relevance, these effects, in particular
those related to quantum (compared to geometrical) capacitances, have been little explored so far~\cite{Gabelli2006Jul}, because they are easily masked by other higher-order effects in experiments. Moreover, temperature-bias-induced screening effects have to our knowledge not been experimentally accessed at all.

\begin{figure}[b!!!]
	\centering
	\includegraphics[width=0.9\columnwidth]{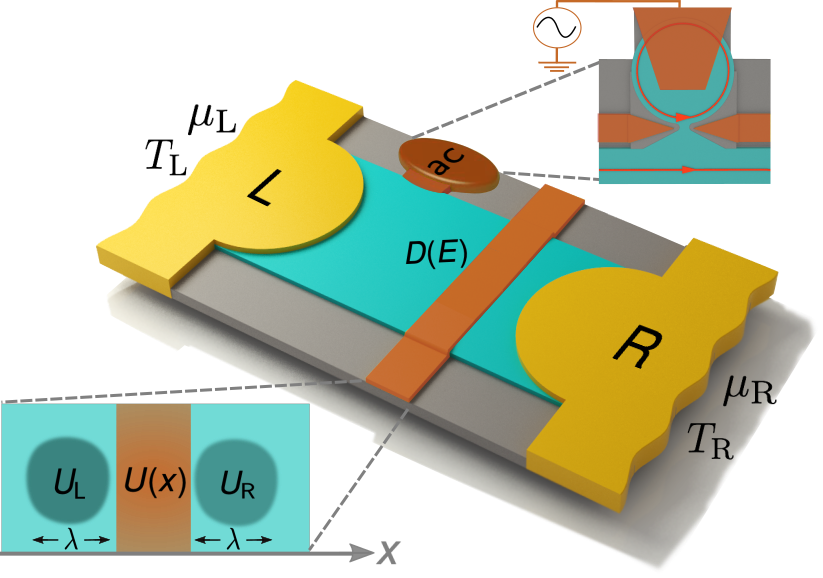}
	\caption{\label{fig_sys}
	Schematic of a coherent mesoscopic conductor, connected to left and right contacts with electrochemical potentials $\mu_\text{L},\mu_\text{R}$ and temperatures $T_\text{L},T_\text{R}$. A third, ac-driven terminal is coupled capacitively only. See the upper inset for the example of a mescocopic capacitor as the ac-source. The coherent conductor has an energy-dependent transmission $D(E)$ (realized, e.g., by a QPC). Lower inset: the potential $U(x)$ creating the energy-dependent scattering region, as well as screening potentials $U_\text{L,R}$ occurring within the screening length $\lambda$ are schematically indicated. 
	}
\end{figure}

In this Letter, we propose a mesoscopic setup that can be exploited to  read out these quantum screening effects.
It consists of a thermally and electrically biased thermoelectric two-terminal conductor, additionally ac-driven by a third local,  capacitively coupled terminal.
The proposed device, as shown in Fig.~\ref{fig_sys}, has an arbitrary energy-dependent transmission, $D(E)$. One possible, simple example for such a conductor could be a quantum point contact (QPC)~\cite{Fertig1987Nov,Buttiker1990Apr,vanHouten1992Mar,Whitney2014Apr,Whitney2015Mar}. 
Away from equilibrium, due to an applied voltage or temperature bias, charge is accumulated at the energy-dependent scatterer, acting as a quantum capacitor. The accumulated charge is screened by charge redistributions at nearby metallic contacts and gates, coupled via geometrical capacitances to the scatterer, and by displacement currents flowing from the contacts.
Treating the electron-electron interactions at a mean-field level~\cite{Buttiker1993Dec}, the result of screening is
a shift of the electrostatic potential
in the conductor, which hence modifies its transmission properties depending on the applied electrical and thermal biases, $D(E)\equiv D(E,\{V,\Delta T\})$~\cite{Christen1996Sep,Pedersen1998Jan,Meair2013Jan,Sanchez2013Jan}.

We find that screening effects due to the stationary biases can already be made visible as corrections to the standard \textit{linear} thermoelectric response of the two-terminal conductor to voltage and temperature biases. These surprising corrections stem from the time-dependent driving, which is locally applied via a third terminal and which could, e.g., be realized by a mesoscopic capacitor~\cite{Buttiker1993Sep,Gabelli2006Jul,Feve07} in the quantum Hall regime~\cite{Fertig1987Nov,Buttiker1990Apr}. 
More specifically, the discovered corrections to the thermoelectric linear-response coefficients are directly proportional to different quantum screening coefficients, which usually only play a role in the nonlinear thermoelectric response of stationary conductors~\cite{Buttiker1993Jun,Buttiker1993Dec,Christen1996Jul,Christen1996Sep,Pedersen1998Jan,Sanchez2013Jan,Meair2013Jan,Texier2018}. In the latter case they occur as higher-order correction effects in the static biases, which are hard to extract from an experiment. In contrast, the correction terms identified here are nonlinear only when accounting for the ac-driving frequency as one of the affinities in a generalized thermoelectric framework~\cite{Ludovico2016Feb}. Screening corrections can hence be switched on and off by adding a local ac-driving and they can thus directly be extracted by comparing standardly detected linear-response coefficients, in the presence and in the absence of the driving.

In the following, we derive charge and heat currents flowing in the time-dependently driven setup shown in Fig.~\ref{fig_sys}, using a Floquet scattering matrix approach~\cite{Buttiker1986Oct,Butcher1990Jun,Blanter2000Sep,Moskalets2011Sep} and carefully considering geometrical and quantum screening effects induced by both voltage and temperature biases. We then elaborate on concrete strategies to exploit the interplay between screening effects and ac-driving in order to read out until now elusive screening coefficients.

\textit{Charge and heat currents in the driven conductor.---}
We consider a coherent mesoscopic conductor connecting two electronic contacts, L and R, via a scattering region with energy-dependent transmission $D(E)$. Here, we assume  contact L to be electrically grounded, i.e. $\mu_\text{L}=\mu_0$, and kept at temperature $T_\text{L}=T_0$, while electrochemical potential and temperature in contact R are assumed to be $\mu_\text{R}=\mu_0+eV$ and $T_\text{R}=T_0-\Delta T$, respectively. Here, $-e$ is the charge of the electron, with $e>0$. In what follows, we set $\mu_0\equiv 0$ as the reference energy.
Furthermore, the conductor is subject to a controlled, local ac-driving applied via a third, capacitively coupled  contact. 

We choose contact R to be the one where the time-averaged charge and energy currents are detected, $I\equiv  I_\text{R}$ and $I^E\equiv  I^E_\text{R}$.
For the conductor shown in Fig.~\ref{fig_sys}, they read~\cite{Blanter2000Sep,Moskalets2011Sep,Nazarov_book}
\begin{subequations}\label{eq_full_curr}
\begin{align}
I&=I_\text{ac}+\frac{e}{h}   \int dE  \: D(E) \: \left\lbrace 
f_\text{R}(E)
-
f_\text{L}(E)
\right\rbrace,  \label{I_R}\\
I^E&=I^{E}_\text{ac}-\frac{1}{h}   \int dE  \:E \: D(E) \:  \left\lbrace 
f_\text{R}(E)
-
f_\text{L}(E)
\right\rbrace, \label{IE_R}
\end{align}
\end{subequations}
with
$f_\alpha(E)=\left[1+\exp\left([E-\mu_\alpha]/k_\text{B}T_\alpha\right)\right]^{-1}$. 
Here, we have split the full currents into a contribution arising from the applied \textit{stationary} temperature and voltage biases (second part of the right hand sides of Eqs.~(\ref{eq_full_curr})) and  contributions,  $I_\text{ac}$ and $I^{E}_\text{ac}$, arising from a time average of the \textit{ac-driving} induced currents.  The latter are given by
\begin{subequations}\label{eq_IJ_s}
\begin{align}
I_\text{ac}&=\frac{e}{h} \: \sum_{n=-\infty}^{\infty}
 \int dE
|S_{n}(E)|^2  \: D(E) \:  \Delta f_0(E_n), \label{eq_I_s} \\
I^{E}_\text{ac}&=-\frac{1}{h} \: \sum_{n=-\infty}^{\infty}
  \int\!dE |S_{n}(E)|^2 D(E) \:E \Delta f_0(E_n). \label{eq_J_s}
\end{align} 
\label{eq_JI_ac}
\end{subequations}
Here, $S_n(E)$ is the $n$-th Fourier component of the scattering matrix of the driven region, see Refs.~\cite{Misiorny2018Feb,Dashti2019Jul} for explicit examples. The function $\Delta f_0(E_n)=f_0(E)-f_0(E_n)$ is a difference between equilibrium Fermi functions, $f_\text{L}(E)\equiv f_0(E)=\left[1+\exp\left(E/k_\text{B}T_0\right)\right]^{-1}$ at energies $E$ and $E_n=E+n\hbar\Omega$, differing by an integer multiple of the ac-driving frequency $\Omega$.
{Eq.~\eqref{eq_JI_ac} relies on the assumption of no backscattering from the conductor towards the source, which is met e.g.~in chiral systems.}
In order to obtain the heat current from the expressions given in Eqs.~\eqref{eq_full_curr}, one needs to evaluate
$
J=I^{E}-VI,
$ and analogous expressions for the separate components of the heat current arising from the stationary biases or ac-driving, alone.

\textit{Linear thermoelectric response to $V$ and $\Delta T$.---}
Starting from the general expressions for charge and energy currents, Eq.~(\ref{eq_full_curr}), we derive expressions for $I$ and $J$ to leading order in the applied biases, $V$ and $\Delta T$, but without expanding in the driving frequency
\begin{equation}\label{eq_lin_resp}
\left(\begin{array}{c} I \\ J \end{array}\right)=\left(\begin{array}{c} I^\text{dir}_\text{ac} \\I_\text{ac}^{E,\text{dir}} \end{array}\right)+\left(\begin{array}{cc} G+G_\text{ac} & L+L_\text{ac} \\ M+M_\text{ac} & K+K_\text{ac} \end{array} \right) \left(\begin{array}{c} V  \\ \Delta T \end{array}\right).
\end{equation}
To obtain this equation, we expand the Fermi functions to linear order in the biases, as well as the energy-dependent transmission probability, which depends on the biases due to screening~\cite{supp,Buttiker1993Jun,Buttiker1993Dec,Christen1996Jul,Christen1996Sep,Pedersen1998Jan,Sanchez2013Jan,Meair2013Jan,Texier2018}
\begin{equation}
D(E,\{V,\Delta T\})= D_0(E)+\frac{1}{2}\frac{d D_0}{dE} \left(\xi eV+  \chi k_\text{B}\Delta T\right). 
\label{DEexp}
\end{equation}
Here, we introduce $D_0(E)=D(E,\{0,0\})$. The coefficient $0\leq \xi \leq 1$ is bounded from above by gauge invariance, while $\chi$ can have any sign and is not bounded.
We evaluate the screening coefficients due to voltage and temperature biases, $\chi$ and $\xi$, for the explicit example of a QPC in the last part of this Letter.  In principle, screening at the QPC of the electrons and holes injected from the local ac-driving should also be accounted for. However, we focus on driving sources operated such that this dynamical, ac-screening effect is negligible~\cite{supp}. 

The first terms appearing in Eq.~\eqref{eq_lin_resp} are to leading order not affected by the applied biases, that is
$I_\text{ac}^\text{dir}=I_\text{ac}\vert_{\Delta T,V=0}$ and $I_\text{ac}^{E,\text{dir}}=I_\text{ac}^E \vert_{\Delta T,V=0}$, with $I_\text{ac}, I_\text{ac}^E$ given in Eqs.~(\ref{eq_I_s}) and (\ref{eq_J_s}). Furthermore, the matrix elements $G,L,M$ and $K$ are the standard, linear response, thermoelectric coefficients
\begin{equation}
    G=\frac{e^2}{h}\mathcal{I}_0, 
    \quad 
    L=\frac{M}{T_0}=-\frac{e}{h} k_\text{B}\mathcal{I}_1, 
    \quad 
    K=\frac{1}{h}  (k^2_\text{B}T_0) \mathcal{I}_2, 
\end{equation}
(see, e.g, Ref.~\cite{Benenti2017Jun} for a review) with
\begin{equation} 
    \mathcal{I}_\ell
    = 
    \int_{-\infty}^{\infty}dE\, 
    D_0(E)   
    \left(\frac{E}{k_\text{B}T_0} \right)^\ell 
    \left(-\frac{\partial
	f_0(E)}{\partial E} \right). 
\end{equation}
Here, $G$ is the electrical conductance, $K$ the thermal one, and $L,M$ the thermoelectric coefficients related to the Seebeck and Peltier coefficients. We emphasize that none of these quantities is affected by the screening effects.  
Of main interest here, are the coefficients $G_\text{ac}, L_\text{ac}, M_\text{ac}$ and $K_\text{ac}$, which modify the standard linear response result and which may, in general, depend nonlinearly on the ac-driving frequency. 
These coefficients all arise from the interplay between the non-equilibrium induced screening effects and the time-dependent driving.
Namely, the applied biases lead to a voltage- and temperature-dependent transmission probability, $D(E)$, which in turn modifies the currents injected due to the local time-dependent driving, when they are scattered at the conductor. The expressions for the coefficients are 
\begin{align}
    G_\text{ac}&=\xi \frac{e^2}{2h}\mathcal{J}_0, & \quad L_\text{ac}&=\chi \frac{k_\text{B}e}{2h} \mathcal{J}_0,\nonumber \\
    M_\text{ac}&=-\xi \frac{k_\text{B}e}{2h} T_0 \mathcal{J}_1, & \quad K_\text{ac}&=-\chi \frac{k^2_\text{B}T_0}{2h}  \mathcal{J}_1, 
\label{eq_coeffs}
\end{align}
where 
\begin{align}
    \label{eq_Jn}
 {\mathcal J}_\ell=
 &\sum_{n}  \int dE   |S_{n}(E)|^2 \frac{dD_0(E)}{dE} \left(\frac{E}{k_\text{B}T_0}\right)^\ell\Delta f_0(E_n).
\end{align}
Interestingly, from Eq.~(\ref{eq_coeffs}) we see that the charge-current and heat-current coefficients are related in a simple way
\begin{equation}\label{eq_coeff_rel}
\frac{G_\text{ac}}{e\xi}=\frac{L_\text{ac}}{k_\text{B}\chi}, \quad \frac{M_\text{ac}}{e\xi}=\frac{K_\text{ac}}{k_\text{B}\chi}\ .
\end{equation}
This derives from the fact that the two pairs of coefficients, $G_\text{ac},M_\text{ac}$ and $L_\text{ac},K_\text{ac}$, respectively stem from voltage- and temperature-induced screening effects.

The relation, Eq.~(\ref{eq_coeff_rel}), demonstrates that the total coefficient matrix in Eq.~(\ref{eq_lin_resp}) does not satisfy Onsager's symmetry relations.  We stress that this breakdown is to be expected, due to the external driving breaking time-reversal symmetry.
Onsager symmetries can be recovered by treating the frequency as an affinity in the adiabatic regime~\cite{Ludovico2016Feb}.

\textit{Weak thermoelectric effect.---}
The origin of the coefficients in Eq.~(\ref{eq_coeffs}) as an interplay between the screening effects and the ac-induced currents becomes formally explicit in the limit of a weak thermoelectric effect, that is, for a conductor with a smooth energy dependence. In this limit, we can expand the transmission probability to first order in energy as $D_0(E)\approx D_0+ED_0'$, where $D_0\equiv D_0(0)$ and $D_0' \equiv dD_0(E)/dE\vert_{E=0}$. Inserting this expansion into the coefficients in Eq.~(\ref{eq_coeffs}) we arrive at
\begin{align}
G_\text{ac}&=-\frac{h}{2e^2}\frac{L\xi}{\mathcal{L}_0T_0} I_{\text{ac},0}^\text{dir}, &  M_\text{ac}&=-\frac{h}{2e^2}\frac{L\xi}{\mathcal{L}_0T_0} J_{\text{ac},0}^\text{dir}
\label{eq_coeffsweak}
\end{align}
and equivalent relations for $L_\text{ac}$ and $K_\text{ac}$ from Eq.~(\ref{eq_coeff_rel}). Here, the thermoelectric coefficient is given by $L=-\frac{e\pi^2}{3h}k_\text{B}^2T_0 D_0'$ in accordance with Mott's law and the Lorenz number is defined as $\mathcal{L}_0=\frac{\pi^2k_\text{B}^2}{3e^2}$. Importantly, the corrections to all response coefficients, $G_\text{ac}$, $L_\text{ac}$, $M_\text{ac}$, and $K_\text{ac}$ become particularly simple in this regime. They are proportional to the screening coefficients $\xi$ and $\chi$, and to the same unperturbed thermoelectric coefficient $L$. Furthermore, they are proportional to the bare charge or heat currents from the time-dependent driving, $I_{\text{ac},0}^\text{dir}$ or $J_{\text{ac},0}^\text{dir}$,  that would flow into contact R for a completely open conductor, $D(E)\rightarrow 1$, and in the absence of stationary biases. Note, however, that  $I_{\text{ac},0}^\text{dir}\equiv 0$, due to the fact that the additional time-dependent driving is local and purely ac. 
This means that in order to obtain non-vanishing corrections to the linear thermoelectric response coefficients of the charge current, $G_\text{ac}$ and $L_\text{ac}$, the energy-dependence of the conductor's transmission probability needs to be at least quadratic. 

\textit{Sensing of quantum screening effects.---}
In typical, purely
statically-biased conductors, the screening effects introduced above occur as higher-order corrections in $\Delta T$ and $V$
~\cite{Christen1996Sep,Pedersen1998Jan,Meair2013Jan,Sanchez2013Jan}, which are hard to clearly identify.
Indeed, screening effects due to a temperature bias have not been observed so far. In the present Letter, we propose to exploit the above introduced interplay  between local ac-driving and quantum screening effects to read out the latter from the modifications of the linear-response coefficients, $G_\text{ac}$, $L_\text{ac}$, $M_\text{ac}$, and $K_\text{ac}$. Note that these are not simply uncontrolled small corrections to  the standard linear-response coefficients, but can be switched on and off at will with the ac-driving.  For the specific readout, we distinguish two situations: (i) the weak thermoelectric case, where at the same time the driving properties are well known, and (ii) the general case of arbitrary $D(E)$, where we do not assume a detailed knowledge of the driving features either.

Case (i) requires the possibility of detecting the heat current response of the conductor. Equations~(\ref{eq_coeff_rel}) and (\ref{eq_coeffsweak}) determine the modifications of the linear-response coefficients.  While $I_{\text{ac},0}^\text{dir}\equiv 0$, one can determine  $J_{\text{ac},0}^\text{dir}$ from a heat current measurement where the ac-driving is applied, but no stationary biases. A stationary charge-current measurement in the presence of a temperature bias yields $L$. With this, one can subsequently directly extract the coefficients $\xi$ and $\chi$ from a detection of $M+M_\text{ac}$ and $K+K_\text{ac}$ compared to $M$ and $K$ in the absence of an ac-driving. 

In case (ii), the functions $\mathcal{J}_0$ and  $\mathcal{J}_1$ are not necessarily known. An experiment could then have two strategies to proceed: either a measurement of all four coefficients, $G_\text{ac}$, $L_\text{ac}$, $M_\text{ac}$, and $K_\text{ac}$, gives access to the four unknown functions $\chi$, $\xi$, $\mathcal{J}_0$ and  $\mathcal{J}_1$, allowing to determine $\chi$ and $\xi$, separately. Otherwise, in an experiment, e.g.~restricted to a measurement of charge-current coefficients only, one could extract the ratio
\begin{equation}
\frac{\chi}{\xi} = \frac{e }{k_\text{B} }\frac{L_\text{ac}}{G_\text{ac}}.
\end{equation}
This would give access to, until now undetected, quantum screening properties due to a thermal bias, as it will be shown in the following  example of a QPC.

\textit{Quantum point contact.---}
As an explicit example, we here consider a scattering region created by a gate-tunable QPC. It can be described by an inverted parabola potential $U(x)=\epsilon-m\omega^2x^2/2$, where $m$ is the effective electron mass, $\omega$ determines the smoothness of the barrier as $\gamma=2\hbar\omega$, and $\epsilon$ is a threshold energy.
The QPC's equilibrium transmission probability is then given by~\cite{Buttiker1990Apr}
\begin{equation}\label{eq_D0}
D_0(E)=\frac{1}{1+\text{exp}\left[-\left(E-\epsilon\right)/\gamma\right]} .
\end{equation}
In order to evaluate screening effects, we follow Refs.~\cite{Pedersen1998Jan,Meair2013Jan} and consider a model of the QPC with two constant potential regions, where the charge is not perfectly screened, one on each side of the QPC, see the inset of Fig.~\ref{fig_sys}. Their size is given by the screening length $\lambda$
We consider a spatially symmetric setup, where the constant-potential regions are equally capacitively coupled to both the QPC split-gate electrodes, with capacitance $C_\text{g}$, and to the respective electronic contact, with capacitance $C$. All other capacitive couplings are assumed to have a negligibly small influence on the screening properties. The detailed derivation of the QPC's scattering properties within a semi-classical, WKB approach is shown in the Supplemental Material~\cite{supp,Connor1968Jan}. 
It yields explicit expressions for the dimensionless coefficients $\chi$ and $\xi$, introduced in Eq.~(\ref{DEexp}), given by
\begin{equation}\label{xichieq}
\xi=\frac{2C+\mathcal{D}}{2C+\mathcal{D}+2C_\text{g}}, \quad \chi=\frac{\mathcal {D}^E}{2C+\mathcal{D}+2C_\text{g}}.
\end{equation}
They contain both the geometric capacitances $C$ and $C_g$, which can be obtained via a careful modeling of the geometry of the actual experimental device (see e.g.~\cite{Deprez2021}), as well as ${\mathcal D}=-e^2\int dE \nu(E) \partial f_0/\partial E$ and ${\mathcal D}^E=-e^2\int dE [E/(k_\text{B}T_0)] \nu(E) \partial f_0/\partial E$, which are due to quantum screening. {In particular, $\mathcal{D}$ is the so-called quantum capacitance~\cite{Gabelli2006Jul,Buttiker1993Sep,Luryi1988,Buttiker1996Sep}, while $\mathcal{D}^E$ (also having units of a capacitance) is related to the charge pile-up in the system due to temperature variations~\cite{Meair2013Jan,Sanchez2013Jan}. They are both quantum properties, as they involve
$\nu(E)$, that} is the total density of states in the two constant-potential regions, $U_\text{L,R}$. The density of states is given by~\cite{Pedersen1998Jan,Wigner1955Apr,Smith1960Apr} 
\begin{align}
    \nu(E)=\frac{1}{\gamma\pi}\times\left\{\begin{array}{ll}
    \mbox{arcosh}\left[\sqrt{\frac{E_\lambda}{\epsilon-E}}\ \right], & \text{for}\  \epsilon-E_{\lambda}<E<\epsilon\\
    \mbox{arsinh}\left[\sqrt{\frac{E_\lambda}{E-\epsilon}}\ \right], & \text{for} \ E>\epsilon \ .
    \end{array}\right. 
      \label{nuE_QPC}
\end{align}
Here, $E_\lambda\equiv m\omega^2\lambda^2/2=\gamma^2 (m \lambda^2/(8\hbar ^2))\equiv\gamma^2/E_\text{box}$, where $E_\text{box}$  and $E_\lambda$ are two energy scales related to the screening length $\lambda$, indicated in the lower inset of Fig.~\ref{fig_sys}.
Importantly, the expression in Eq.~(\ref{xichieq}) clearly shows that the factor $\chi/\xi=\mathcal{D}^E/\left(2C+\mathcal{D}\right)$, that is most easily accessible by the above described readout scheme (ii), gives access to  quantum screening properties due to a thermal bias, encoded in $\mathcal{D}^E$.

\begin{figure}[t]
	\centering
	\includegraphics[scale=1]{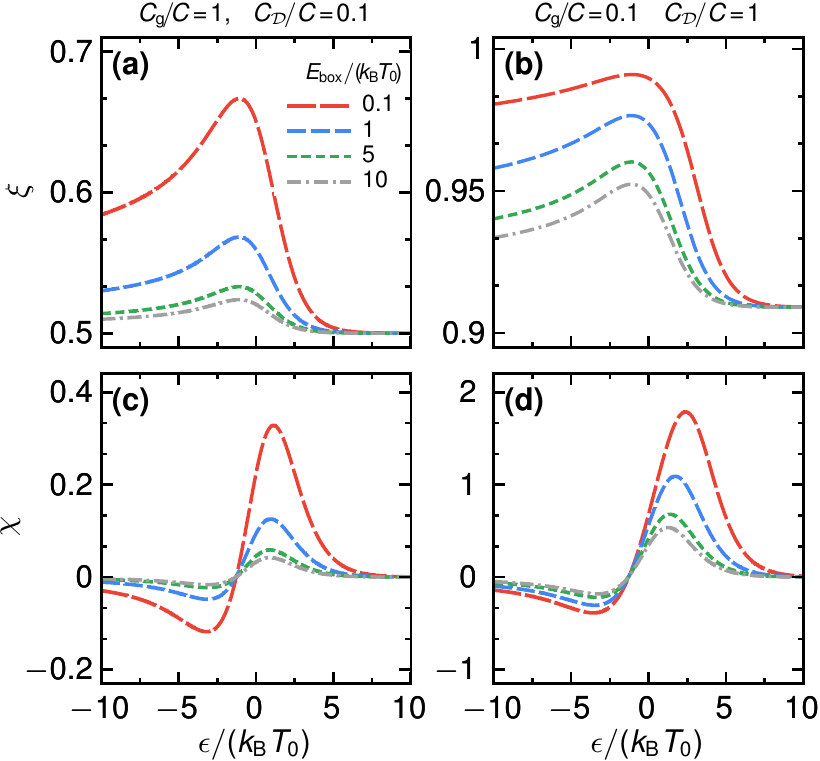}
	\caption{\label{fig_xichi}
	Coefficients $\xi$, in (a), (b), and $\chi$, in (c), (d) as a function of $\epsilon/(k_\text{B}T_0)$ for a set of different $E_\text{box}/(k_\text{B}T_0)$, see legend in (a), and for different values of the parameters $C_\text{g}/C$ and $C_{\mathcal D}/C$. Here, the smoothness of the barrier is $\gamma/(k_\text{B}T_0)=0.1$.}
\end{figure}
Conveniently, $\xi$ and $\chi$ can be expressed~\cite{supp} in terms of three dimensionless energy parameters  $\epsilon/(k_\text{B}T_0)$, $E_\text{box}/(k_\text{B}T_0)$, and $\gamma/(k_\text{B}T_0)$, as well as in terms of two dimensionless, capacitive parameters $C_\text{g}/C,$ and  $C_{\mathcal D}/C$, where $C_{\mathcal D}=e^2/(8\pi  k_\text{B}T_0) $. The combination $C_\mathcal{D}\gamma/E_\text{box}$, containing the screening length $\lambda$, gives the typical magnitude of the quantum
capacitances ${\mathcal D}$ and ${\mathcal D}^E$~\cite{supp}.
{From Eq.~(\ref{xichieq}) it follows that in the limit of dominant
capacitive
coupling to the gate, $C_\text{g} \gg C, \mathcal D, \mathcal{D}^E$ both coefficients are small, $\xi,\chi \ll 1$, leading to a tiny modification of the transmission with respect to $D_0(E)$. 
In the regime $C \gg C_g, \mathcal{D},\mathcal{D}^E$, with dominant capacitive coupling to the contacts, $\xi\to 1$ and the internal potentials $U_\text{L,R}$ are shifted by the same amount as the electrochemical potentials $\mu_{\text{L,R}}$. However $\chi \ll 1$, i.e., the effect of temperature is small. For the regime of dominating quantum capacitances, $\mathcal{D},\mathcal{D}^E \gg C,C_g$, both coefficients $\chi,\xi$ can be of order one.
Taken together, the effect on the transmission due to applied bias or temperature is determined by the relation between quantum and geometric capacitances. 
Note that both the magnitude of the quantum capacitance, resulting from imperfect screening at the QPC, as well as the classical, capacitive couplings between different parts of the conductor are affected by the strength of electron-electron interactions.
} 

In Fig.~\ref{fig_xichi}, we plot both $\xi$ and $\chi$ as a function of $\epsilon$ for a representative set of parameters. The dependence on $\epsilon$ in these plots is entirely due to quantum capacitances, which, unlike geometric ones, {depend on the transmission properties of the conductor.} We see that $\xi$ shows a qualitatively similar behavior in both panels, Figs.~$\ref{fig_xichi}$~(a) and (b), with a maximum around $\epsilon=0$, approaching $C/(C+C_\text{g})$ for $\epsilon/(k_\text{B}T_0) \rightarrow \infty$ and decaying slowly with increasingly negative $\epsilon$. The magnitude of the variations with $\epsilon$ is however larger for $C_\text{g} \gg C_{\mathcal D}$. The trend is opposite for $\chi$, in Figs.~\ref{fig_xichi}~(c) and (d), where larger variations with $\epsilon$ occur for $C_{\mathcal D} \gg C_\text{g}$. Overall, $\chi$ shows a qualitatively similar, alternating-sign behavior in both panels, with a negative peak at $\epsilon<0$ and a positive peak at $\epsilon>0$, both of the order of $k_\text{B}T_0$ away from the origin. For $\epsilon/(k_\text{B}T_0) \rightarrow \pm \infty$, $\chi$ approaches zero.

\textit{Conclusions.---}
We have shown how the interplay between a local ac-driving and quantum screening effects due to stationary thermal and electrical biases impacts the standard, stationary linear response of a thermoelectric conductor. We use this to put forward a proposal for the direct readout of -- until now elusive -- quantum screening effects, from tunable modifications of linear-response coefficients.  We expect presently available experimental techniques~\cite{Bauerle2018Apr,Brun2019,Dubois2012b,Fletcher2013Nov,Johnson2017} to allow for the proposed readout of quantum screening effects. For the same parameters as in Fig.~\ref{fig_xichi} and considering as a probe a mesoscopic capacitor with driving frequency around $1\,$GHz and escape time of $30\,$ps~\cite{Mahe2010} and a static voltage bias of $10\,\mu$V, we get a rough estimate of $5\,$pA and $1\,$fW for the corrections to the charge and heat currents, respectively. These values are increased by reducing $\gamma$ (i.e.~sharpening the energy-dependence of the QPC transmission). The findings of such an experiment could test predictions for screening coefficients, as they are shown in Fig.~\ref{fig_xichi}. 

We furthermore foresee that with this very same setup, known screening effects can be used to perform tomography on single-electron sources (attached to the third contact in our scheme)~\cite{Fletcher2013Nov}. Also, the controlled modification of thermoelectric response coefficients by the driving is expected to be of interest for the improvement of heat engines~\cite{Zhou2015Oct}.

\begin{acknowledgments}
We thank F.~Taddei, R.~S.~Whitney, G.~Haack and S.~Lara-Avila for interesting discussions and G.~F\`eve and D.~S\'anchez for useful comments on the manuscript. Funding from the Knut and Alice Wallenberg Foundation through the Academy Fellow program (J.S., N.D., M.A. and M.M.), from the European Union's H2020
research and innovation program under grant agreement No 862683 UltraFastNano (M.A. and J.S.), and from the Swedish VR are gratefully acknowledged.
\end{acknowledgments}


%

\end{document}


\title{Supplemental material: Readout of quantum screening effects}
	
	\author{Nastaran Dashti}
	\affiliation{Department of Microtechnology and Nanoscience (MC2), Chalmers University of Technology, S-412 96 G\"oteborg, Sweden}
	
	\author{Matteo Acciai}
	\affiliation{Department of Microtechnology and Nanoscience (MC2), Chalmers University of Technology, S-412 96 G\"oteborg, Sweden}

	\author{Sara Kheradsoud}
	\affiliation{Department of Physics, Lund University, S-221 00 Lund, Sweden}
	
	\author{Maciej Misiorny}
	\affiliation{Department of Microtechnology and Nanoscience (MC2), Chalmers University of Technology, S-412 96 G\"oteborg, Sweden}

	\author{Peter Samuelsson}
	\affiliation{Department of Physics, Lund University, S-221 00 Lund, Sweden}
		
	\author{Janine Splettstoesser}
	\affiliation{Department of Microtechnology and Nanoscience (MC2), Chalmers University of Technology, S-412 96 G\"oteborg, Sweden}
	
	\date{\today}
		\maketitle

\section{Screening effects in weakly nonlinear response}\label{app_screening}

In the following, we provide a detailed discussion of the effect of screening in the weakly non-linear transport regime. Various aspects of the result have been presented in different papers over several decades, see e.g.  Refs.~\cite{Buttiker1993Jun,Buttiker1993Dec,Christen1996Jul,Christen1996Sep,Pedersen1998Jan,Sanchez2013Jan,Meair2013Jan}. However, it is our impression that a complete, self-consistent discussion is missing. Since the material is mainly known, but still is of central interest to our work, we present it in detail in this section of the supplemental material.
\subsection{QPC potential, screening regions and semi-classical approach}
\begin{figure}[t]
  \centering
  {\includegraphics[scale=0.25]{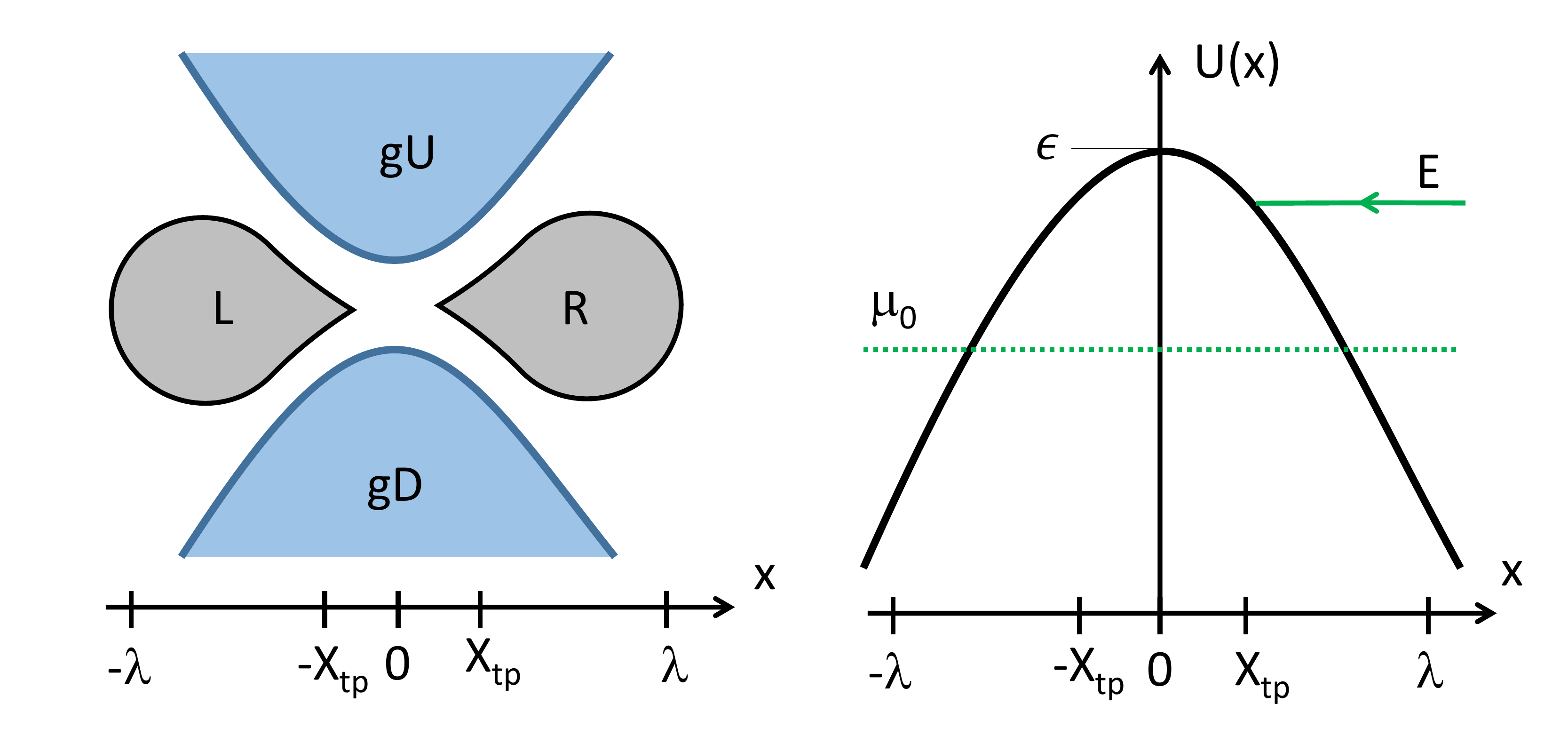}}
  \caption{(a) Schematic top-view of QPC, showing regions L and R where the charge is not screened. The upper (gU) and lower (gD) split gate electrodes are also shown. (b) Energy and potential sketch. At the QPC, the electrostatic potential can be approximated by an inverted parabola, with top energy $\epsilon$. The width of the potential parabola determines the smoothness $\gamma$ of the barrier. For further details see the text.}
\label{fig_app_sys}
\end{figure}

In Fig.~\ref{fig_app_sys}, left panel, we show a schematic top-view of the QPC region sketched in { Fig.~1 in the main text}. We assume that there is only one conduction mode open in the QPC and that the problem hence is effectively 1D, along the $x$-axis. Indicated in the figure are two regions, L and R, on each side of the QPC midpoint at $x=0$. In these two regions, of the size of the screening length $\lambda$, it is assumed that the charge is not completely screened.  The electrostatic potential $U(x)$ of the QPC is taken to be an inverted parabola~\cite{Fertig1987Nov,Buttiker1990Apr}, see right panel of Fig.~\ref{fig_app_sys}, with 
\begin{equation}
eU(x)=\epsilon-\frac{m\omega^2}{2} x^2\ ,
\end{equation}
where $\epsilon$ determines the top of the potential, at $x=0$. Here, $m$ is the effective mass of the electron and $\hbar \omega/2=\gamma$, where $\gamma$ is the smoothness of the transmission probability, {given in Eq.~12 in the main text}. 

Let's now consider an electron incident from e.g. the right at an energy $E$, where the energy is counted from $\mu_0\equiv0$, the electrochemical potential of the reservoirs at equilibrium. This is shown in panel~(b) of Fig.~\ref{fig_app_sys}. Using a semiclassical, WKB analysis, the electron has a classical turning point at a position $x_\text{tp}=x_\text{tp}(E)$, obtained  from  $E=\epsilon-m \omega^2x_\text{tp}^2/2$, that is 
\begin{equation}
x_\text{tp}=\sqrt{\frac{2(\epsilon-E)}{m\omega^2}}.
\end{equation}
As a consequence we can say that the regions L and R, where charges are not fully screened, are defined by
\begin{equation}
-\lambda<x<-x_\text{tp}, \quad x_\text{tp}<x<\lambda
\end{equation}
respectively, as shown in Fig.~\ref{fig_app_sys}. We note that the expression for $x_\text{tp}$ formally holds only for $E<\epsilon$. For $E>\epsilon$, the result would be non-physical (imaginary) and we instead take $x_\text{tp}=0$, that is, there is no classical turning point and region L and R are in direct contact. Moreover, for sufficiently low energies $E_\text{min}$, the turning point $x_\text{tp}(E)$ reaches the boundaries of the non-perfectly screened region. This happens when $x_\text{tp}=\lambda$, which gives $E_\text{min}=\epsilon-E_\lambda$, where we introduced for later convenience $E_\lambda=m\omega^2\lambda^2/2$.   

\subsection{Scattering matrix, semiclassical approach}
To find the scattering matrix $S$ for the QPC, we point out that the length of the scattering region is taken to be $-\lambda<x<\lambda$. We first note that, quite generally, the scattering matrix for the QPC can be written as 
\begin{equation}
S=\left( \begin{array}{cc} ie^{i\phi(E)}\sqrt{1-D_0(E)} & e^{i\phi(E)}\sqrt{D_0(E)}  \\  e^{i\phi(E)}\sqrt{D_0(E)} & ie^{i\phi(E)}\sqrt{1-D_0(E)}  \end{array} \right)\ ,
\end{equation}
where we take into account that the QPC is spatially symmetric and impose the unitarity condition for $S$, i.e. $S^{\dagger}S=1$. The transmission probability $D_0(E)$ is given by {Eq.~(12) in the main text:}
\begin{equation}
 D_0(E)=\frac{1}{1+e^{-(E-\epsilon)/\gamma}}.
\label{eq_app_D}
\end{equation}
The scattering phase $\phi(E)$ is the dynamical phase acquired when traversing the QPC. Starting with the case $E_\text{min}<E<\epsilon$, it is obtained by integrating the semiclassical, position-dependent momentum $p(x)=\sqrt{2m[E-U(x)]}$ over the path through region L and R,
\begin{equation}
\phi(E)=\frac{\sqrt{2m}}{\hbar}\left(\int_{-\lambda}^{-x_\text{tp}}dx+\int^{\lambda}_{x_\text{tp}}dx\right) \sqrt{E-\epsilon+\frac{m\omega^2}{2}x^2}
\end{equation} 
giving
\begin{align}
\phi(E)&=\frac{m\omega x_\text{tp}^2}{\hbar}\times\\
& \left[ \frac{\lambda}{x_\text{tp}}\sqrt{\left(\frac{\lambda}{x_\text{tp}}\right)^2-1}-\ln \left( \frac{\lambda}{x_\text{tp}}+\sqrt{\left(\frac{\lambda}{x_\text{tp}}\right)^2-1}\right)\right] \nonumber \\
&=\frac{\epsilon-E}{\gamma}\times\nonumber\\
&\left[ \sqrt{\frac{E_\lambda}{\epsilon-E}\left(\frac{E_\lambda}{\epsilon-E}-1\right)}- \mbox{arcosh}\left(\sqrt{\frac{E_\lambda}{\epsilon-E}}\right) \right]\ ,\nonumber
 \end{align} 
where we used that $m\omega x_\text{tp}^2/\hbar=(\epsilon-E)/\gamma$.

\begin{figure}[t]
  \centering
  {\includegraphics[scale=0.28]{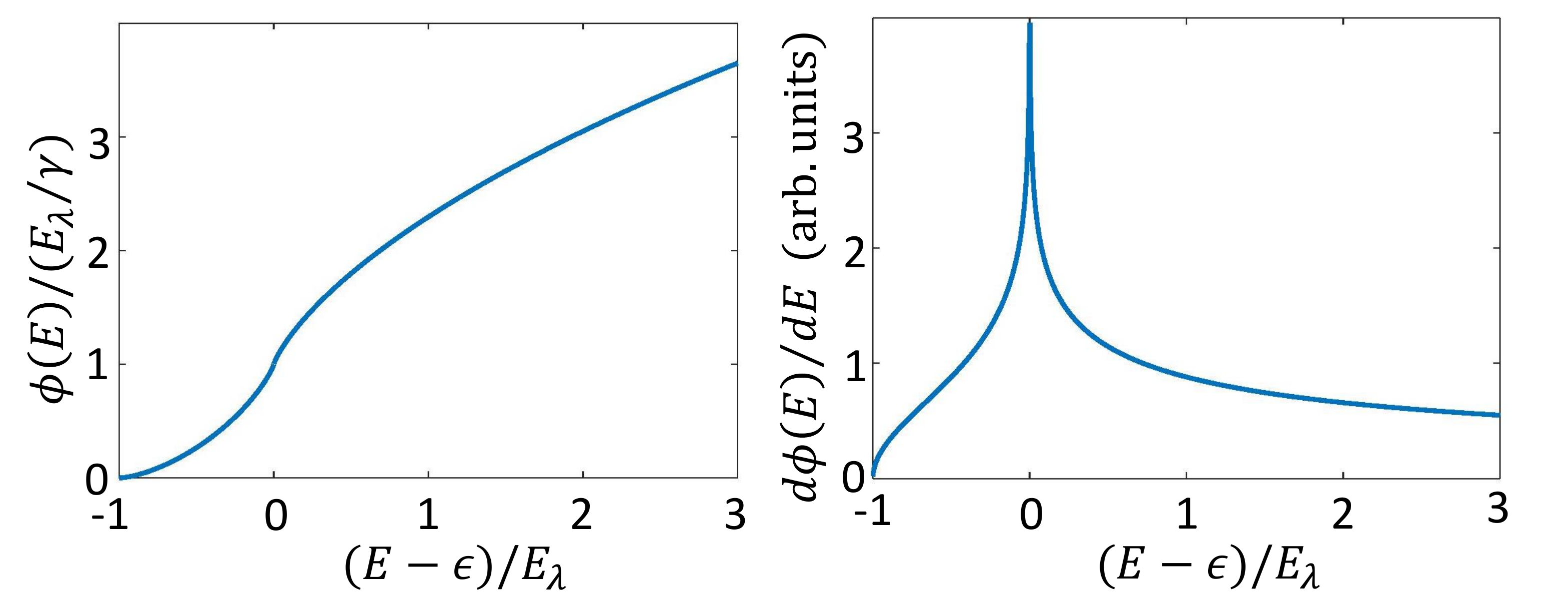}}
  \caption{(a) Normalized scattering phase as a function of energy. (b) Energy derivative of the scattering phase, proportional to the density of states (see text).}
\label{fig_phase}
\end{figure}

For energies $E<E_\text{min}$ the acquired phase is zero. For energies above the potential top, $E>\epsilon$, $x_\text{tp}=0$ and we can proceed as above and write the acquired phase
\begin{align}
\phi(E)&=\frac{1}{\hbar}\int_{-\lambda}^{\lambda} dx \sqrt{2m\left(E-\epsilon+\frac{m\omega^2}{2}x^2\right)}
\end{align} 
giving
\begin{align}
\phi(E)&=\frac{E-\epsilon}{\gamma}\times\\
&\left[\sqrt{\frac{E_\lambda}{E-\epsilon}\left(\frac{E_\lambda}{E-\epsilon}+1\right)}+\mbox{arsinh}\left(\sqrt{\frac{E_\lambda}{E-\epsilon}}\right) \right]\ .\nonumber
\end{align} 
We note that the phase, in addition to the amplitudes, depends on the energy scale $E_\lambda$. In Fig.~\ref{fig_phase}, the normalized phase $\phi(E)/(E_\lambda/\gamma)$, is plotted as a function of energy $(E-\epsilon)/E_\lambda$. It is clear that the phase has a cusp at $E=\epsilon$.

We stress that it is in principle possible to perform a full quantum mechanical calculation of the scattering matrix elements, following Refs.~\cite{Connor1968Jan,Fertig1987Nov}. Since the main interest here is to get a qualitative picture of the physics, we however judge that a semiclassical treatment is sufficient.

\subsection{Density of states and injectivities} 
The next step is to consider the density of states (DOS). It is known~\cite{Wigner1955Apr,Smith1960Apr} that the global DOS, $\nu(E)$, of an arbitrary scatterer is related to the scattering matrix $S$ as 
\begin{equation}
\nu(E)=\frac{1}{2\pi i}\mbox{tr}\left[S^{\dagger}\frac{dS}{dE}\right]=\frac{1}{\pi}\frac{d\phi(E)}{dE}\ .
\end{equation} 
From the expressions for $\phi(E)$ above, we have, for $E_\text{min}<E<\epsilon$,
\begin{equation}
\nu(E)=\frac{1}{\gamma \pi} \mbox{arcosh}\left(\sqrt{\frac{E_\lambda}{\epsilon-E}}\right) 
\label{nuE1}
\end{equation} 
and for $E>\epsilon$ we have
\begin{equation}
\nu(E)=\frac{1}{\gamma \pi} \mbox{arsinh}\left(\sqrt{\frac{E_\lambda}{E-\epsilon}}\right), 
\label{nuE2}
\end{equation} 
in line with Ref.~\cite{Pedersen1998Jan}; {see also Eq.~(14) in the main text}. In Fig.~\ref{fig_phase}, we plot the energy derivative of the phase. It is clear that the phase derivative has a singularity at $E=\epsilon$, a consequence of the semiclassical approximation. As is clear below, this singularity is integrable, that is, it does not prevent an analysis of the energy integrated DOS, entering the final result.  

\begin{figure}[t]
  \centering
  {\includegraphics[scale=0.35]{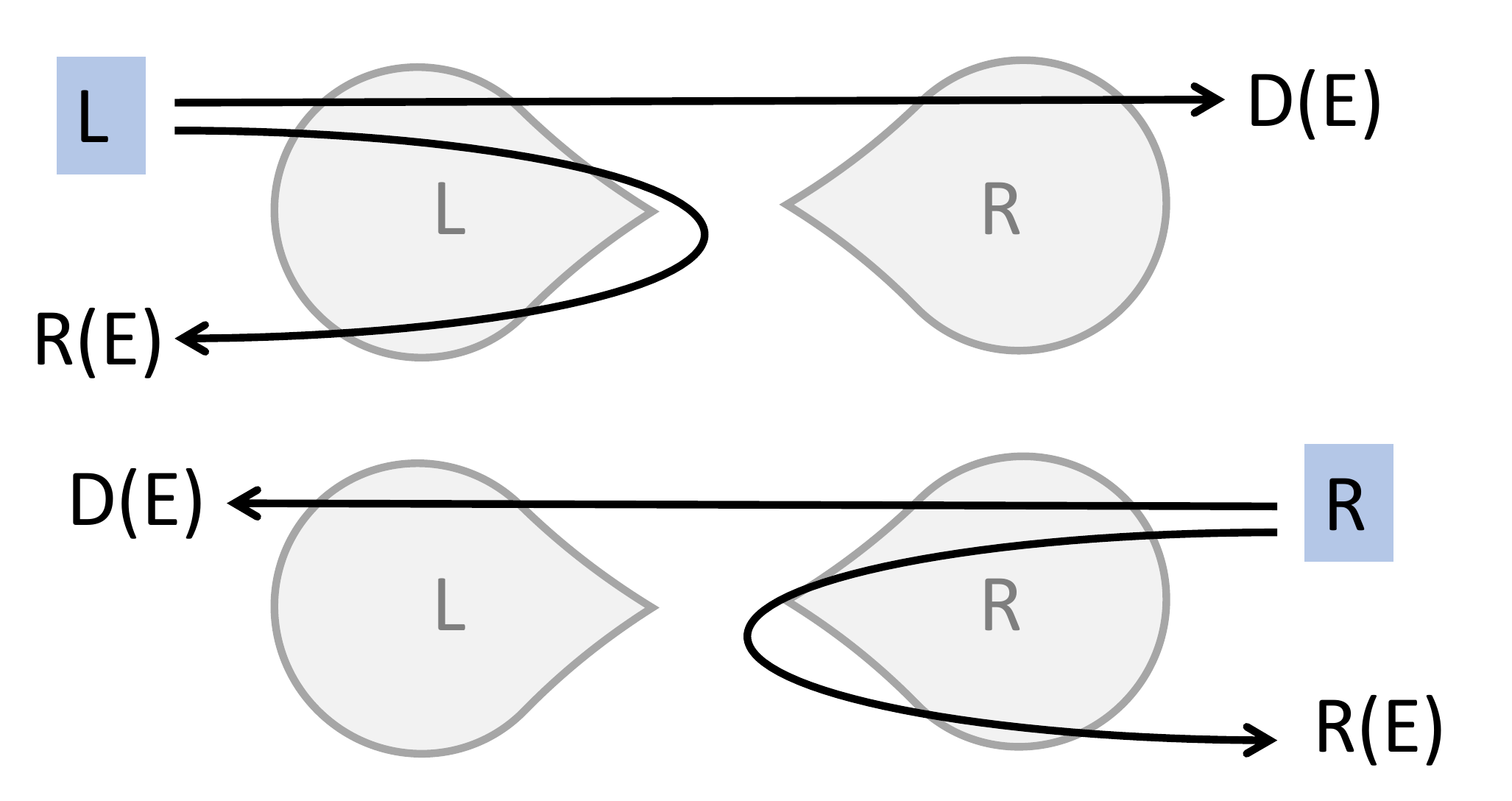}}
  \caption{Schematic of scattering paths contribution to the density of states.}
\label{fig_paths}
\end{figure}

We note that since the QPC is symmetric, half of the states are on each side of the saddle point, such that
\begin{equation}
\nu_\text{L}(E)=\nu_\text{R}(E)=\nu(E)/2\ .
\end{equation} 
Here $\nu_\alpha(E)$ is thus the local density of states in the regions $\alpha=\text{L,R}$, see Fig.~\ref{fig_app_sys}. Based on the local density of states, we can follow the discussion in Ref.~\cite{Christen1996Jul} to calculate the partial density of states and the related injectivities for the QPC. To this aim, it is helpful to consider the trajectories for incoming particles from the left and right, shown in Fig.~\ref{fig_paths}.
%
From these paths we can write down the local, partial density of states $\nu_{\alpha\beta\delta}$, where $\delta=\text{L,R}$  denotes the reservoir from which a particle is incident on the scatterer,  $\alpha=\text{L,R}$ denotes the reservoir to which the particle is emitted from the scatterer, and $\beta=\text{L,R}$ denotes the region of the density of states to which the path contributes. This gives, by inspection, writing out all eight cases explicitly, 
\begin{eqnarray}
 \nu_\text{LLR}(E)&=&\frac{1}{2}D(E)\nu_\text{L}(E), \quad \nu_\text{LLL}(E)=R(E)\nu_\text{L}(E), \nonumber \\
\nu_\text{RLL}(E)&=&\frac{1}{2}D(E)\nu_\text{L}(E),  \quad \nu_\text{RLR}(E)=0,  \nonumber \\
 \nu_\text{LRR}(E)&=&\frac{1}{2}D(E)\nu_\text{R}(E), \quad \nu_\text{LRL}(E)=0,  \nonumber \\
\nu_\text{RRL}(E)&=&\frac{1}{2}D(E)\nu_\text{R}(E), \quad \nu_\text{RRR}(E)=R(E)\nu_\text{R}(E), \nonumber\\
\end{eqnarray} 
Here $D(E)$ and $R(E)=1-D(E)$ are the probabilities for the different paths to occur, given that one particle is incident from the reservoir. The factor $1/2$ in front of the terms with $D(E)$ tells that the particle only traverses the region in one direction (out of two possible), thus contributing to only one half of the total DOS. We stress that the following relation holds 
\begin{eqnarray}
 \sum_{\alpha,\beta,\delta}\nu_{\alpha \beta \delta}(E)=\nu_\text L(E)+\nu_\text R(E)=\nu(E).
\end{eqnarray} 
From the partial, local DOS we can construct the injectivities $\nu_{\beta\delta}(E)$ by summing over the reservoirs to which the particle is emitted. Explicitly, we have
\begin{eqnarray}
 \nu_\text{LL}(E)&=&\nu_\text{LLL}(E)+\nu_\text{RLL}(E)=\frac{1}{2}[1+R(E)]\nu_\text{L}(E)\nonumber \\
  \nu_\text{RR}(E)&=&\nu_\text{LRR}(E)+\nu_\text{RRR}(E)=\frac{1}{2}[1+R(E)]\nu_\text{R}(E)
  \nonumber \\
\nu_\text{RL}(E)&=&\nu_\text{LRL}(E)+\nu_\text{RRL}(E)= \frac{1}{2}D(E)\nu_\text{R}(E) \nonumber \\
\nu_\text{LR}(E)&=&\nu_\text{LLR}(E)+\nu_\text{RLR}(E)= \frac{1}{2}D(E)\nu_\text{L}(E).\nonumber \\
\end{eqnarray} 
In the same way, one can obtain the \textit{emissivities} of the QPC, however, as they are not needed for this calculation, we do not present them here. 

\subsection{Induced charge, bare and screened} 
As a result of the applied potential and temperature biases, $V_\alpha$ and $\Delta T_\alpha$ at the reservoirs $\alpha=\text{L,R}$, charge is injected into the QPC regions. First, the bare charges $Q_\text{L}^\text{(b)}$ and $Q_\text{R}^\text{(b)}$ on the two QPC regions can be written in terms of the injectivities as 
\begin{equation}\label{eq_totalscreeningcharge}
\begin{split}
Q_\text{L}^\text{(b)}&=\mathcal{D}_\text{LL} V_\text{L}+\mathcal{D}_\text{LR} V_\text{R}+\mathcal{D}_\text{LL}^E \Delta T_\text{L}+\mathcal{D}_\text{LR}^E \Delta T_\text{R},  \\
Q_\text{R}^\text{(b)}&=\mathcal{D}_\text{RL} V_\text{L}+\mathcal{D}_\text{RR} V_\text{R}+\mathcal{D}_\text{RL}^E \Delta T_\text{L}+\mathcal{D}_\text{RR}^E \Delta T_\text{R}, 
\end{split}
\end{equation} 
Here we have introduced the total, energy integrated charge~\cite{Buttiker1993Jun} and entropic~\cite{Sanchez2013Jan} injectivities
\begin{eqnarray}
\mathcal{D}_{\alpha\beta} & = & -e^2\int dE \nu_{\alpha\beta}(E)\frac{df_0}{dE}\\ 
\mathcal{D}_{\alpha\beta}^E &=& -e\int dE \frac{E}{T_0}\nu_{\alpha\beta}(E)\frac{df_0}{dE}.
\end{eqnarray} 
Note that the total charge injectivities are given with the units of capacitance. 

As a result of the injected charge, the system responds by trying to screen it. In the QPC regions, the electrostatic potentials are shifted $U_\text{L}$ and $U_\text{R}$ away from their equilibrium values and screening charges $Q_\text{L}^\text{(s)}$ and $Q_\text{R}^\text{(s)}$ are induced. Following the same semiclassical approach as for the scattering matrix~\cite{Christen1996Jul}, we can write the screening charges as
\begin{eqnarray}
Q_\text{L}^\text{(s)}=-\mathcal{D}_\text{L}U_\text{L}, \quad Q_\text{R}^\text{(s)}=-\mathcal{D}_\text{R}U_\text{R}\ ,
\end{eqnarray} 
where we introduced the energy integrated, local density of states 
\begin{eqnarray}
\mathcal{D}_{\alpha}=-e^2\int dE \nu_{\alpha}(E)\frac{df_0}{dE}, \quad \mathcal{D}_\text L=\mathcal{D}_\text R=\frac{\mathcal{D}}{2}.
\label{enintdensofstates}
\end{eqnarray} 
Here ${\mathcal D}$ is the total, energy integrated DOS in the system (in the units of capacitance). 

\begin{figure}[h]
  \includegraphics[scale=0.28]{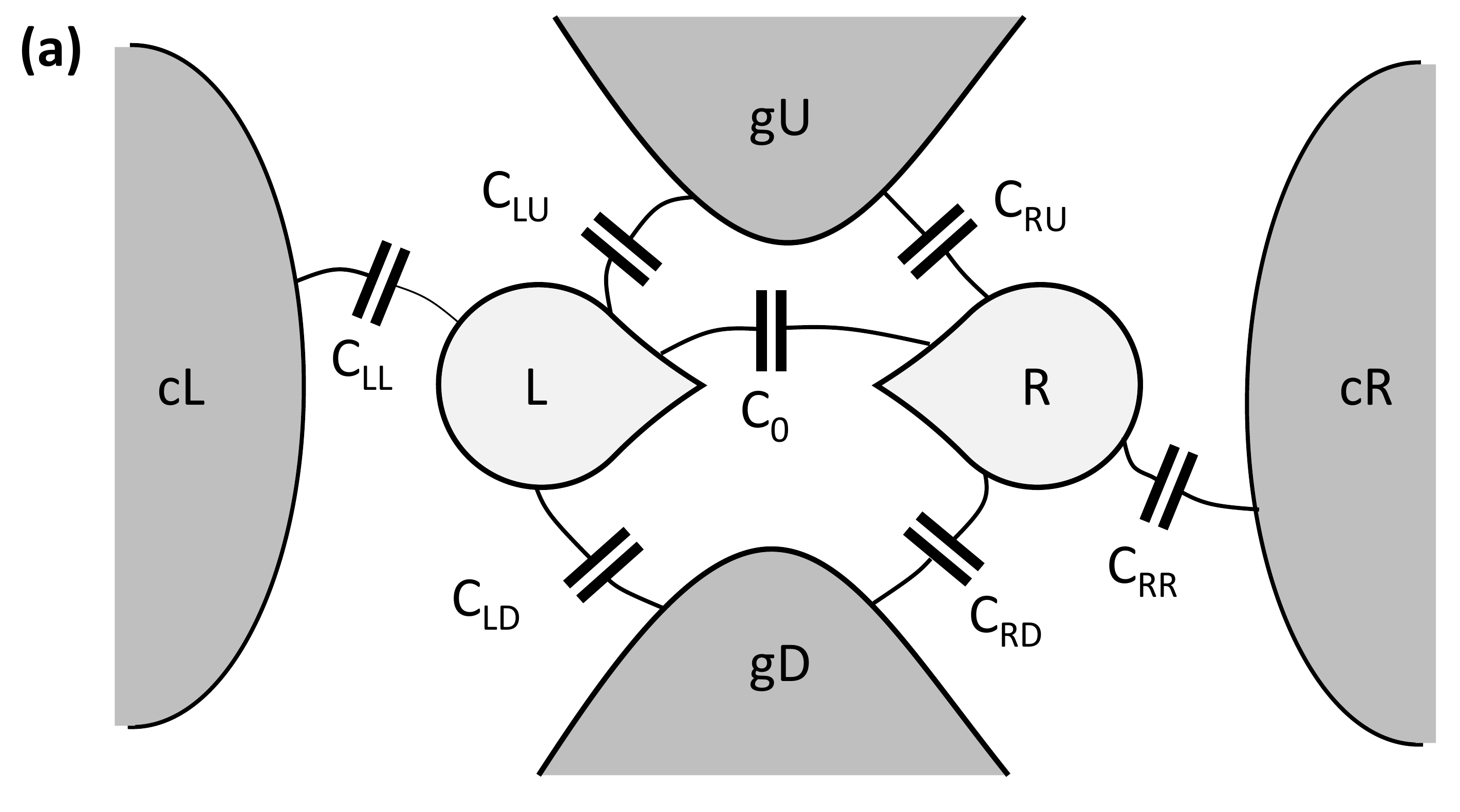}\\
  \includegraphics[scale=0.28]{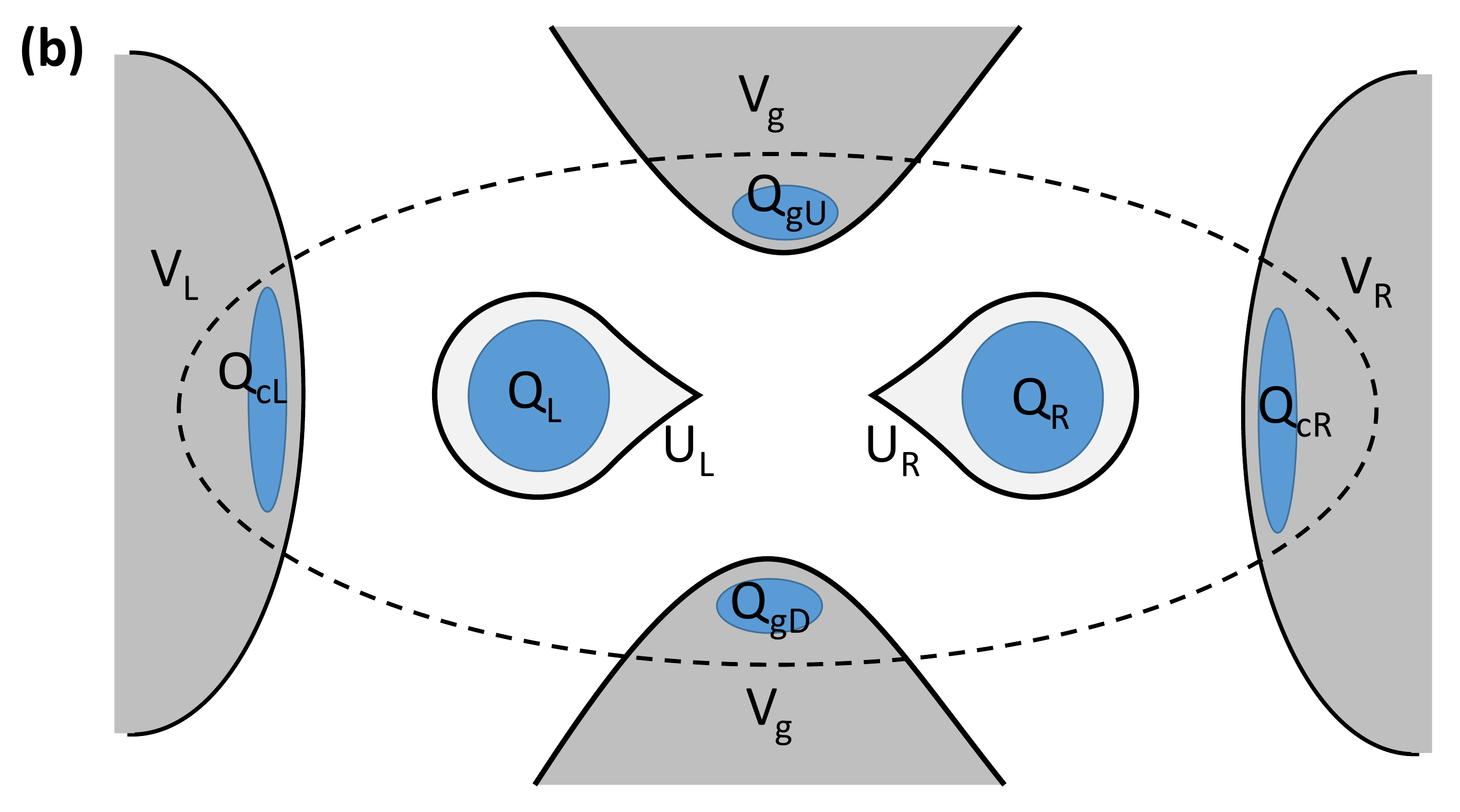}
  \caption{(a) Most relevant geometric capacitances in the system. (b) Induced charges, in QPC and on nearby metallic gates and reservoirs. The Gauss region, inside which the total charge is zero, is shown with dashed lines.}
\label{figcharge}
\end{figure}

The total induced charges in the two regions is then given by the sums of bare and screened charges, $Q_\text L=Q_\text L^{(\text b)}+Q_\text L^{(\text s)}$ and $Q_\text R=Q_\text R^{(\text b)}+Q_\text R^{(\text s)}$, giving
\begin{eqnarray}
Q_\text L&=&{\mathcal D}_{\text{LL}} V_\text L+{\mathcal D}_{\text{LR}} V_\text R+{\mathcal D}_{\text{LL}}^E \Delta T_\text L+{\mathcal D}_{\text{LR}}^E \Delta T_\text R-\frac{{\mathcal D}}{2}U_\text L, \nonumber \\
Q_\text R&=&{\mathcal D}_{\text{RL}} V_L+{\mathcal D}_{\text{RR}} V_\text R+{\mathcal D}_{\text{RL}}^E \Delta T_\text L+{\mathcal D}_{\text{RR}}^E \Delta T_\text R-\frac{{\mathcal D}}{2}U_\text R. \nonumber \\
&\quad &
\label{Qeq1}
\end{eqnarray} 
As a next step, we take into account that the total charges $Q_\text L$ and $Q_\text R$ also couple capacitively to nearby metallic gates and reservoirs, as well as to each other. For the QPC system, the most relevant capacitive couplings are shown in {Fig.~\ref{figcharge}~(a)}. As a result of the capacitive interactions, shown schematically in {Fig.~\ref{figcharge}~(b)}, there will be charges induced on the surfaces of the metallic gates and reservoirs, such that inside a Gauss region~\cite{Christen1996Sep} the total charge is zero.  

We can thus write the electrostatic relations 
\begin{eqnarray}
Q_ \text L&=& C_{\text{LL}}(U_\text L-V_\text L)+C_{\text{LU}}(U_\text L-V_\text g) + C_{\text{LD}}(U_\text L-V_\text g) \nonumber \\
&+& C_0(U_\text L-U_\text R) \nonumber \\
Q_\text R&=& C_{\text{RR}}(U_\text R-V_\text R)+C_{\text{RU}}(U_\text R-V_\text g)+ C_{\text{RD}}(U_\text R-V_\text g) \nonumber \\
&+& C_0(U_\text R-U_\text L),
\label{Qeq2}
\end{eqnarray} 
where we have assumed that the same potential $V_\text g$ is applied to both gate electrodes (as is normally the case for a split gate). We can now combine the expressions for the charge in Eqs.~(\ref{Qeq1}) and (\ref{Qeq2}), giving relations for the induced potentials $U_\text L$ and $U_\text R$ in terms of the applied voltages $V_\text L,V_\text R$, temperatures $\Delta T_\text L, \Delta T_\text R$ and the gate voltage $V_\text g$, in a matrix form as
\begin{eqnarray}\label{eq:U}
\left(\begin{array}{c} U_\text L \\ U_\text R \end{array} \right)&=&\left(\begin{array}{cc} \xi_{\text{LL}}  & \xi_{\text{LR}} \\  \xi_{\text{RL}}  & \xi_{\text{RR}} \end{array} \right)\left(\begin{array}{c} V_\text L \\ V_\text R \end{array} \right)+\left(\begin{array}{c} v_\text L \\ v_\text R \end{array} \right)V_\text g \nonumber \\
&+& \dfrac{k_\text B}{e}\left(\begin{array}{cc} \chi_{\text{LL}}  & \chi_{\text{LR}} \\  \chi_{\text{RL}}  & \chi_{\text{RR}} \end{array} \right)\left(\begin{array}{c} \Delta T_\text L \\ \Delta T_\text R \end{array} \right).
\end{eqnarray} 
The coefficients $\xi_{\alpha\beta}, \chi_{\alpha\beta}$ and $v_{\alpha}$ are the characteristic potentials we need for the further evaluation. An explicit calculation gives for the voltage ones
\begin{eqnarray}
\xi_{\text{LL}}&=&\frac{1}{Z} [(2 C_\text R + 2 C_{\text{RR}} + {\mathcal D}) (C_{\text{LL}} + {\mathcal D}_{\text{LL}}) \nonumber \\
&+& 2C_0 (C_{\text{LL}} + {\mathcal D}_{\text{LL}} + {\mathcal D}_{\text{RL}})] \nonumber \\
\xi_{\text{RL}}&=&\frac{1}{Z} [(2 C_\text L + 2 C_{\text{LL}} + {\mathcal D}) {\mathcal D}_{\text{RL}}  \nonumber \\
&+& 2 C_0 (C_{\text{LL}} + {\mathcal D}_{\text{LL}} + {\mathcal D}_{\text{RL}})] \nonumber \\
\xi_{\text{RR}}&=&\frac{1}{Z} [2 C_\text L + 2 C_{\text{LL}} + {\mathcal D}) (C_{\text{RR}} + D_{\text{RR}} \nonumber \\
&+& 2 C_0 (C_{\text{RR}} +{\mathcal  D}_{\text{LR}}+ {\mathcal D}_{\text{RR}})]  \nonumber \\
\xi_{\text{LR}}&=&\frac{1}{Z} [(2 C_\text R + 2C_{\text{RR}} + {\mathcal D}) {\mathcal D}_{\text{LR}} \nonumber \\
&+& 2 C_0(C_{\text{RR}}+ {\mathcal D}_{\text{LR}}+ {\mathcal D}_{\text{RR}})],
\label{ucoeffs}
\end{eqnarray}
where we introduced $C_\text L=C_{\text{LD}}+C_{\text{LU}}, C_\text R=C_{\text{RD}}+C_{\text{RU}}$ and the denominator 
\begin{eqnarray}\label{eq:z}
Z&=&2 C_0 (C_\text L + C_{\text{LL}} + C_\text R + C_{\text{RR}} +{\mathcal D}) \nonumber \\
&+& \frac{1}{2}(2 C_\text L + 2 C_{\text{LL}} +{\mathcal D}) (2 C_\text R + 2 C_{\text{RR}} + {\mathcal D})
\end{eqnarray}
For the temperature ones we get
\begin{eqnarray}
\chi_{\text{LL}}&=&\frac{e}{k_\text{B}}\frac{(2 C_\text R + 2 C_{\text{RR}} + {\mathcal D}) {\mathcal D}_{\text{LL}}^E + 2 C_0 ({\mathcal D}_{\text{LL}}^E + {\mathcal D}_{\text{RL}}^E)}{Z} \nonumber \\
\chi_{\text{RL}}&=&\frac{e}{k_\text{B}}\frac{(2 C_\text L + 2 C_{\text{LL}} + {\mathcal D}) {\mathcal D}_ {\text{RL}}^E + 2 C_0 ({\mathcal D}_{\text{LL}}^E + {\mathcal D}_{\text{RL}}^E)}{Z} \nonumber \\
\chi_{\text{RR}}&=&\frac{e}{k_\text{B}}\frac{(2 C_\text L + 2 C_{\text{LL}} +{\mathcal D}) {\mathcal D}_{\text{RR}}^E + 2 C_0 ({\mathcal D}_{\text{LR}}^E+ {\mathcal D}_{\text{RR}}^E)}{Z}  \nonumber \\
\chi_{\text{LR}}&=&\frac{e}{k_\text{B}}\frac{(2 C_\text R + 2 C_{\text{RR}} + {\mathcal D}) {\mathcal D}_{\text{LR}}^E + 2 C_0 ({\mathcal D}_{\text{LR}}^E + {\mathcal D}_{\text{RR}}^E)}{Z} \nonumber \\
\end{eqnarray}
and for the gate potential ones
\begin{eqnarray}
v_{\text L}&=&\frac{2 C_0 (C_\text L + C_\text R) + C_\text L (2 C_\text R + 2 C_{\text{RR}} + {\mathcal D})}{Z} \nonumber \\
v_{\text R}&=&\frac{2 C_0 (C_\text L + C_\text R) + C_\text R (2 C_\text L + 2 C_{\text{LL}} + {\mathcal D})}{Z}. 
\label{vcoeffs}
\end{eqnarray}
We point out that in the limit considered by Meair and Jacquod~\cite{Meair2013Jan}, our result reduces to theirs.

\subsection{Transport quantities, weak non-linear expansion}

The electrical and energy/heat currents both depend on the transmission probability $D(E)$. Away from equilibrium, in the presence of electrical and/or thermal bias, the transmission probability becomes dependent on the applied biases $V_\text L,V_\text R$ and $\Delta T_\text L, \Delta T_\text R$. The equilibrium value of the gate potential is used to regulate the barrier top energy $\epsilon$ and the width, determining $\gamma$. Throughout the discussion we keep the gate potential constant, at its equilibrium value, i.e. $V_\text g=0$. As discussed above, the applied biases affect the scattering properties by modifying the potentials $U_\text L,U_\text R$, that is, we can write
\begin{equation}
D(E)\equiv D(E,U_\text L[\{V_\alpha,\Delta T_\alpha\}],U_\text R[\{V_\alpha,\Delta T_\alpha\}]),
\end{equation}
where $\{V_\alpha,\Delta T_\alpha\}=V_\text L,V_\text R,\Delta T_\text L,\Delta T_\text R$. Within the weakly non-linear approximation we expand $D(E)$ to leading order in the biases, as 
\begin{eqnarray}
D(E)&\equiv & D_0(E)+\frac{\partial D(E)}{\partial U_\text L}\left(\frac{\partial U_\text L}{\partial V_\text L}V_\text L+ \frac{\partial U_\text L}{\partial V_\text R}V_\text R \right. \nonumber \\
&+& \left.  \frac{\partial U_\text L}{\partial \Delta T_\text R}\Delta T_\text R+\frac{\partial U_\text L}{\partial \Delta T_\text L}\Delta T_\text L\right) +   \frac{\partial D(E)}{\partial U_R}\left(\frac{\partial U_R}{\partial V_L}V_L \right.   \nonumber \\
&+& \left. \frac{\partial U_\text R}{\partial V_\text R}V_\text R+\frac{\partial U_\text R}{\partial \Delta T_\text R}\Delta T_\text R+\frac{\partial U_\text R}{\partial \Delta T_\text L}\Delta T_\text L\right),
\end{eqnarray}
where $D_0(E)$ is the equilibrium transmission probability {in Eq.~(12) in the main text} and all partial derivatives are evaluated at $\{V_\alpha,\Delta T_\alpha\}=0$. Making use of the characteristic potentials we can write, collecting the bias terms,
\begin{eqnarray}
D(E)&\equiv & D_0(E)+\left(\frac{\partial D(E)}{\partial U_\text L}\xi_{\text{LL}}+\frac{\partial D(E)}{\partial U_\text R}\xi_{\text{RL}}\right)V_\text L \nonumber \\
&+& \left(\frac{\partial D(E)}{\partial U_\text L}\xi_{\text{RL}}+\frac{\partial D(E)}{\partial U_\text{R}}\xi_{\text{RR}}\right)V_\text R \nonumber \\
&+&\left(\frac{\partial D(E)}{\partial U_\text L}\chi_{\text{LL}}+\frac{\partial D(E)}{\partial U_\text R}\chi_{\text{RL}}\right)\dfrac{k_\text B \Delta T_\text L}{e} \nonumber \\
&+&\left(\frac{\partial D(E)}{\partial U_\text L}\chi_{\text{RL}}+\frac{\partial D(E)}{\partial U_\text{R}}\chi_{\text{RR}}\right)\dfrac{k_\text B \Delta T_\text R}{e}.
\end{eqnarray}
Now, it can be shown that gauge invariance guarantees ~\cite{Buttiker1993Dec,Christen1996Sep} the relation
\begin{equation}
\frac{\partial D(E)}{\partial V_\text L}+\frac{\partial D(E)}{\partial V_\text R}+\frac{\partial D(E)}{\partial V_\text g}+e\frac{\partial D(E)}{\partial E}=0.
\end{equation}
Written in terms of the characteristic potentials we have
\begin{eqnarray}
&&\frac{\partial D(E)}{\partial U_\text L}\left(\xi_{\text{LL}}+\xi_{\text{LR}}+v_{\text L}\right)+\frac{\partial D(E)}{\partial U_\text R}\left(\xi_{\text{RL}}+\xi_{\text{RR}}+v_{\text R}\right) \nonumber \\
&=&-e\frac{\partial D(E)}{\partial E}.
\end{eqnarray}
Then, using the condition that the sum of the characteristic potential at a given region is unity, gives
\begin{equation}
\xi_{\text{LL}}+\xi_{\text{LR}}+v_{\text L}=1, \quad \xi_{\text{RL}}+\xi_{\text{RR}}+v_{\text R}=1
\end{equation}
and hence
\begin{equation}
\frac{\partial D(E)}{\partial U_\text L}+\frac{\partial D(E)}{\partial U_\text R}=-e\frac{\partial D(E)}{\partial E}.
\end{equation}
Following our assumption that the QPC scattering potential is symmetric around $x=0$ we can write 
\begin{equation}
\frac{\partial D(E)}{\partial U_\text L}=\frac{\partial D(E)}{\partial U_\text R}=-\frac{e}{2}\frac{\partial D(E)}{\partial E}.
\end{equation}
Inserting this into the expression for $D(E)$ and recalling that partial derivatives are evaluated at $\{V_\alpha,\Delta T_\alpha\}=0$, we arrive at ($V_\text g=0$)
\begin{eqnarray}
D(E)&\equiv & D_0(E)- \frac{e}{2}\frac{dD_0(E)}{dE}(\xi_\text L V_\text L+ \xi_\text R V_\text R \nonumber \\
&+&\chi_\text L \dfrac{k_\text B \Delta T_\text L}{e}+\chi_\text R \dfrac{k_\text B \Delta T_\text R}{e}),
\end{eqnarray}
where we introduced, for shortness, $\xi_\text L=\xi_{\text{LL}}+\xi_{\text{RL}}, \xi_\text R=\xi_{\text{RR}}+\xi_{\text{LR}}, \chi_\text L=\chi_{\text{LL}}+\chi_{\text{RL}}$ and $\chi_\text R=\chi_{\text{RR}}+\chi_{\text{LR}}$.

\subsection{Symmetric setup}
For the completely symmetric capacitive situation considered in the main text, we have $C_\text L=C_\text R \equiv C_\text g$, $C_{\text{LL}}=C_{\text{RR}}\equiv C$. As is also clear from the discussion above, we can write the DOS expressions ${\mathcal D}_{\text{LR}}={\mathcal D}_{\text{RL}}, \mathcal{D}_{\text{RR}}={\mathcal D}_{\text{LL}}$ and ${\mathcal D}_{\text{LR}}^E={\mathcal D}_{\text{RL}}^E, {\mathcal D}_{\text{RR}}^E={\mathcal D}_{\text{LL}}^E$. This together allows us to write the relevant characteristic potentials
\begin{eqnarray}
\xi_\text L&=&\xi_\text R=\frac{2C+{\mathcal D}}{2C+{\mathcal D}+2C_\text g}, \quad \chi_\text L=\chi_\text R= \frac{e{\mathcal D}^E/k_\text{B}}{2C+{\mathcal D}+2C_\text g},  \nonumber \\
\end{eqnarray}
noting that ${\mathcal D}/2={\mathcal D}_{\text{LL}}+{\mathcal D}_{\text{LR}}$ and ${\mathcal D}^E/2={\mathcal D}_{\text{LL}}^E+{\mathcal D}_{\text{LR}}^E$. We note that, due to the symmetric setup, neither $\xi_\text L,\xi_\text R$ nor $\chi_\text L,\chi_\text R$ are dependent on $C_0$. Performing a rescaling ${\mathcal D}^E \rightarrow  (k_\text{B}/e) {\mathcal D}^E$ and putting $\xi \equiv \xi_\text L=\xi_\text R, \chi \equiv \chi_\text L= \chi_\text R$, we arrive at {Eq.~(13) in the main text}.

Making use of {Eq.~(2) in the main text}, we can then directly write down the linear response (in voltage and temperature) modifications of the charge and energy currents due to the {ac-driven terminal}. The biasing arrangements are discussed in the main text: $V_\text{L}=0, V_\text{R}=-V,\Delta T_\text{L}=0, \Delta T_\text{R}=-\Delta T$. {Following the notation of the main text, we write the currents as $I=I_\text{ac}^\text{dir}+GV+L\Delta T+\delta I_\text{ac}$ and $I^E=I_\text{ac}^{E,\text{dir}}+MV+K\Delta T+\delta I_\text{ac}^E$}, where
\begin{eqnarray}
\delta I_\text{ac}&=&\frac{e^2}{2h}\left(\xi V+\frac{k_\text B\chi}{e} \Delta T \right) \nonumber \\
&\times & \sum_{n=-\infty}^{\infty} \int dE \ |S_n(E)|^2 \frac{d D_0(E)}{d E} \left[f_0(E)-f_0(E_n)\right], \nonumber \\
\delta I_\text{ac}^E&=&-\frac{e}{2h}\left(\xi V+\frac{k_\text B\chi}{e} \Delta T \right)  \nonumber \\
&\times& \sum_{n=-\infty}^{\infty} \int dE \ |S_n(E)|^2  \frac{d D_0(E)}{d E} E\left[f_0(E)-f_0(E_n)\right] .\nonumber \\
\end{eqnarray}
From these expressions we directly arrive at the expressions for $G_\text{ac}, L_\text{ac}, M_\text{ac}$ and $K_\text{ac}$ in
{Eq.~(7) in the main text}.

\subsection{Weak thermoelectric effect}
It is particularly interesting to investigate the case with a weak thermoelectric effect, resulting from a weakly energy dependent transmission probability, expanded as
\begin{eqnarray}
D_0(E)&=&D_0+E\left. \frac{d D_0}{d E}\right \vert_{E=E_0}+\frac{E^2}{2}\left. \frac{d^2 D_0}{d E^2}\right \vert_{E=E_0}.... \nonumber \\
&\equiv&  D_0+ED_0'+\frac{E^2}{2}D_0''...
\end{eqnarray}
Keeping only leading order in energy dependence, we have the well known~\cite{Butcher1990Jun} linear response coefficients
\begin{equation}
G=\frac{e^2}{h}D_0, \,\, L=\frac{M}{T_0}=-\frac{eT_0}{h}\frac{(\pi k_\text{B})^2}{3}D_0', \,\, K={\mathcal L}_0 T_0 G_0
\end{equation}
where ${\mathcal L}_0=(\pi k_\text{B})^2/(3e^2)$ is the Lorenz number.  The direct source currents become (to leading order) independent on the thermoelectric effect, as
\begin{equation}
I_\text{ac}^{\text{dir}}=D_0 I_{\text{ac},0}^{\text{dir}}, \quad I_\text{ac}^{E,\text{dir}}=D_0 I_{\text{ac},0}^{E,\text{dir}},
\end{equation}
where $I_{\text{ac},0}^{\text{dir}}$ and $I_{\text{ac} ,0}^{E,\text{dir}}$ are the bare charge and energy currents of the source. For the source-dependent linear response terms, given by {Eq.~(7) in the main text}, we get for the conductance
\begin{eqnarray}
G_\text{ac}&=&\xi \frac{e^2}{2h}\sum_{n=-\infty}^{\infty}\int dE \ |S_n(E)|^2 D'_0 \left[f_0(E)-f_0(E_n)\right] \nonumber \\
&=& \frac{e}{2} \xi D'_0 I_{\text{ac},0}^{\text{dir}}= -\frac{h}{2e^2 T_0 {\mathcal L}_0}\xi L I_{\text{ac},0}^{\text{dir}}.
\end{eqnarray}
That is, $G_\text{ac}$ is proportional to the bare source current $I_{\text{ac},0}^{\text{dir}}$, the thermoelectric coefficient $L$, and the screening characteristic potential $\xi$. For the thermoelectric coefficient $L_\text{ac}$ we get, in the same way,
\begin{eqnarray}
L_\text{ac} =-\frac{hk_\text{B}}{2e^3 T_0 {\mathcal L}_0}\chi L I_{\text{ac},0}^{\text{dir}},
\end{eqnarray}
proportional to $G_\text{ac}$. For the energy current terms we have
\begin{eqnarray}
M_\text{ac}&=&-\xi \frac{e}{2h}\sum_{n=-\infty}^{\infty}\int dE\ |S_n(E)|^2 D_0' E\left[f_0(E)-f_0(E_n)\right] \nonumber \\
&=&\frac{e}{2}\xi D_0' I_{\text{ac},0}^{E,\text{dir}} = -\frac{h}{2e^2 T_0 {\mathcal L}_0}\xi L I_{\text{{ac}},0}^{E,\text{dir}}
\end{eqnarray}
and, in the same way,
\begin{eqnarray}
K_\text{ac} =-\frac{hk_\text{B}}{2e^3 T_0 {\mathcal L}_0}\chi L I_{\text{{ac}},0}^{E,\text{dir}}.
\end{eqnarray}
The expressions for $G_\text{ac}$ and $M_\text{ac}$ are the ones given in {Eq.~(10) in the main text}.

\section{Screening due to ac driving}\label{app:screening_time}
 
 In general, not only the injected charges due to stationary voltage and temperature biases lead to screening potentials induced on the QPC, but also the time-dependently injected current from the side-coupled source results in screening effects. In this section we show how one should proceed to include these effects; at the same time we demonstrate {that under relevant operating regimes of the ac source} these effects actually play a minor role.
 
 In this work, we focus on the regime of weakly nonlinear effects. Therefore the effect of temperature and voltage biases enters as separate summands in Eq.~(\ref{eq_totalscreeningcharge}). For the charge accumulated on the barriers due to the time-dependent driving, an additional term enters here, which is proportional to the excitation due to the driving. In the particular driving protocol chosen in the main paper, this extra term enters the expression for $Q_\text{L}^{(\text{b})}$ and contains a factor $\Omega$.
 
 This leads to a first relevant estimate. In the limit $\Omega\ll eV_\alpha/\hbar,k_\text{B}T_\alpha/\hbar$, the contribution to the screening charge due to the driving can be expected to be small. Note however, that this also leads to small corrections to the response coefficients {[Eq.~(7) in the main text] and therefore a more detailed analysis is needed to neglect the screening effects related to the driving. In the following we present the analysis for the charge current only, as an identical reasoning can be repeated for the energy one.}
 
 We introduce the spectral function
 \begin{equation}
     i(E)=\sum_{n}|S_n(E)|^2\Delta f_0(E_n)
 \end{equation}
 which gives the nonequilibrium energy distribution of electrons injected by the source.
 For instance, considering an adiabatically driven mesoscopic capacitor~\cite{Gabelli2006Jul} as the source, one has~\cite{Dashti2019Jul}
 \begin{equation}
     |S_n(E)|^2=(2\Omega\sigma)^2e^{-2|n|\Omega\sigma},
     \label{eq:mes-cap-spectral}
 \end{equation}
 where $\sigma$ is the time width of the pulse. Note, however, that this expression is only for illustration purposes and this specific shape is not required to prove our point. With the spectral function, one can obtain the time-averaged current $\bar{I}$ over a long time as $\bar{I}=-eh^{-1}\int dE\, i(E)$. Since we are interested in a stationary regime where time-average currents are measured, we can then obtain the average injected charges due to the ac drive as
 \begin{equation}
     Q_\alpha^\text{(b,ac)}=-e\int dE\,\nu_{\alpha\text{L}}(E)i(E)\,.
 \end{equation}
 This provides an extra term to Eq.~\eqref{eq_totalscreeningcharge}, which, in turn, modifies Eq.~\eqref{Qeq1}, eventually leading to extra contributions $U'_\alpha$ to the screening potentials in Eq.~\eqref{eq:U}. These extra terms can be expressed as
 \begin{equation}
     U'_\alpha=\sum_\beta\frac{\zeta_{\alpha\beta}}{\mathcal{D}_{\alpha\beta}}Q_\beta^{(\text{b,ac})}
 \end{equation}
 The quantities $\zeta_{\alpha\beta}$ are dimensionless coefficients that express the system response to the injected charges due to the ac drive. Explicitly, they read
 \begin{align}
     \zeta_{\alpha\alpha}&=\frac{\mathcal{D}_{\alpha\alpha}}{Z}(C_{\alpha\alpha}+C_\alpha+C_0+\mathcal{D}/2),\\
     \zeta_{\alpha\beta}&=\frac{\mathcal{D}_{\alpha\beta}}{Z}C_0\quad(\text{for }\alpha\neq\beta),
 \end{align}
 with $Z$ given in Eq.~\eqref{eq:z}. Defining $\zeta_\text{L}=\zeta_\text{LL}+\zeta_\text{RL}$ and $\zeta_\text{R}=\zeta_\text{LR}+\zeta_\text{RR}$, for the symmetric setup discussed in the main text one finds
 \begin{equation}
     \zeta_\text{L}=\zeta_\text{R}=\frac{\mathcal{D}}{2C+\mathcal{D}+2C_\text{g}}\equiv{\zeta}.
 \end{equation}
 Gathering all the results, we eventually find the following expression for the charge current $I=(G+G_\text{ac}+\widetilde{G})V+(L+L_\text{ac}+\widetilde{L})\Delta T+I_\text{ac}^\text{dir}$. Here, $G_\text{ac}$ and $L_\text{ac}$ are the corrections to the standard thermoelectric response discussed in the main text. The coefficients $\widetilde{G}$ and $\widetilde{L}$ are the new contribution resulting from taking into account screening effects due to the ac source itself too. Explicitly, they read
 \begin{equation}
 \begin{split}
     \widetilde{G}&=-\frac{e^4\zeta}{2h\mathcal{D}}\int dE\nu(E)i(E)\!\int dE\frac{dD_0}{dE}\left(-\frac{\partial f_0}{\partial E}\right),\\
     \widetilde{L}&=-\frac{e^3\zeta}{2hT_0\mathcal{D}}\int dE\nu(E)i(E)\!\int\! dE E\frac{dD_0}{dE}\left(-\frac{\partial f_0}{\partial E}\right).
\end{split}
 \end{equation}
 As discussed in the main text, it is possible to isolate the contributions to the current uniquely due to the source, thus having access to $G_\text{ac}+\widetilde{G}$ and $L_\text{ac}+\widetilde{L}$. It is then clear that if there is a regime where $\widetilde{G}\ll G_\text{ac}$ and $\widetilde{L}\ll L_\text{ac}$, the screening effects due to the ac source can be neglected, leading to the discussion in the main text. To show that such a condition exists, we have considered the case of an adiabatic mesoscopic capacitor, see Eq.~\eqref{eq:mes-cap-spectral}. An explicit evaluation of the extra terms $\widetilde{G}$ and $\widetilde{L}$ with this source shows that they are indeed negligible in the regime $\hbar/\sigma\ll|\epsilon|$.

\section{Key expressions}\label{app:integrals}
In the symmetric case considered in the main text, the numerical evaluation of the screening coefficients $\xi$ and $\chi$ amounts to evaluating the total, energy integrated charge and entropic injectivities, $\mathcal D$ and $\mathcal D^E$. Here we present some details of this evaluation. From Eq.~\eqref{enintdensofstates} we have
\begin{eqnarray}
{\mathcal D}&=&\frac{e^2}{4\pi k_\text{B}T_0\gamma}\int_{\epsilon-E_{\lambda}}^{\epsilon}dE \frac{\mbox{arcosh}\left(\sqrt{E_\lambda/(\epsilon-E)}\right)}{\cosh^2(E/[2k_\text{B}T_0])} \nonumber \\
&+&\frac{e^2}{4\pi k_\text{B}T_0\gamma} \int_{\epsilon}^{\infty}dE \frac{\mbox{arsinh}\left(\sqrt{E_\lambda/(E-\epsilon)}\right)}{\cosh^2(E/[2k_\text{B}T_0])} .
\end{eqnarray}
Changing variables as $x=(\epsilon-E)/E_\lambda$ and $y=(E-\epsilon)/E_\lambda$, we get
\begin{eqnarray}
{\mathcal D}&=&\frac{e^2E_\lambda}{4\pi \gamma k_\text{B}T_0}\left[\int_{0}^{1}dx~\mbox{arcosh}\left(\sqrt{\frac{1}{x}}\right) \frac{1}{\cosh^2([\epsilon_0-x \epsilon_\lambda]/2)} \right. \nonumber \\
&+& \left. \int_{0}^{\infty}dy~\mbox{arsinh}\left(\sqrt{\frac{1}{y}}\right) \frac{1}{\cosh^2([\epsilon_0+y \epsilon_\lambda]/2)} \right],
\end{eqnarray}
where we introduced the dimensionless energies $\epsilon_\lambda=E_\lambda/[k_\text{B}T_0]$ and $\epsilon_0=\epsilon/[k_\text{B}T_0]$. To have a more shorthand notation we first write 
\begin{equation}
{\mathcal D}=2{\mathcal D}_0\left({\mathcal F}_0+{\mathcal G}_0\right), \quad {\mathcal D}_0=\frac{e^2E_\lambda}{8\pi \gamma k_\text{B}T_0}, 
\end{equation}
where the dimensionless integrals are
\begin{eqnarray}
{\mathcal F}_n&=&\int_{0}^{1}dx~x^n\mbox{arcosh}\left(\sqrt{\frac{1}{x}}\right) \frac{1}{\cosh^2([\epsilon_0-x \epsilon_\lambda]/2)} \nonumber \\
{\mathcal G}_n&=&\int_{0}^{\infty}dy~y^n\mbox{arsinh}\left(\sqrt{\frac{1}{y}}\right) \frac{1}{\cosh^2([\epsilon_0+y \epsilon_\lambda]/2)} . \nonumber \\
\end{eqnarray}
In this notation, convenient for the numerics, we can write the characteristic potential 
\begin{equation}
\xi=\dfrac{1+\frac{{\mathcal D}}{2C}}{1+\frac{{\mathcal D}}{2C}+\frac{C_g}{C}}=\frac{1+c_d({\mathcal F}_0+{\mathcal G}_0)}{1+c_g+c_d({\mathcal F}_0+{\mathcal G}_0)}, 
\label{xisymm}
\end{equation}
introducing yet another shorthand notation with dimensionless quantities $c_d={\mathcal D}_0/C, c_g=C_g/C$. This form shows clearly the different, independent dimensionless parameters that controls $\xi$, namely $\epsilon_\lambda,\epsilon_0,c_g,c_d$.  

In the same way we have
\begin{eqnarray}
{\mathcal D}^E&=&\frac{e^2}{4\pi  \gamma k_\text{B}^2T_0}\int_{\epsilon-E_{\lambda}}^{\epsilon}dE \frac{E}{T_0} \frac{\mbox{arcosh}\left(\sqrt{E_\lambda/(\epsilon-E)}\right)}{\cosh^2(E/[2k_\text{B} T_0])} \nonumber \\
&+& \frac{e^2}{4\pi  \gamma k_\text{B}^2T_0} \int_{\epsilon}^{\infty}dE\frac{E}{T_0} \frac{\mbox{arsinh}\left(\sqrt{E_\lambda/(E-\epsilon)}\right)}{\cosh ^2(E/[2k_\text{B}T_0])}.   \nonumber \\
\end{eqnarray}
Making the same variable substitutions as for ${\mathcal D}$ we have
\begin{equation}
{\mathcal D}^E=2{\mathcal D}_0\left[\epsilon_0\left({\mathcal F}_0+{\mathcal G}_0\right)+\epsilon_\lambda \left(-{\mathcal F}_1+{\mathcal G}_1\right)\right].
\end{equation}
We can thus write the relevant, dimensionless characteristic potential function
\begin{equation}
\chi=\frac{{\mathcal D}^E}{2C+\mathcal{D}+2C_g}=\frac{c_d\left[\epsilon_0\left({\mathcal F}_0+{\mathcal G}_0\right)+\epsilon_\lambda \left(-{\mathcal F}_1+{\mathcal G}_1\right)\right]}{1+c_g+c_d({\mathcal F}_0+{\mathcal G}_0)}.
\label{chisymm}
\end{equation}
We note that since the integrals ${\mathcal F}_n$ and ${\mathcal G}_n$  are functions of $\epsilon_0$ and $\epsilon_\lambda$, the same four parameters $\epsilon_0,\epsilon_\lambda,c_d$ and $c_g$ determine both $\xi$ and $e\chi /k_\text{B}$. 

\section{Screening coefficients, limiting expressions}\label{app:limiting}

In the main text explicit expression for the screening coefficients are given in the limiting cases $\epsilon_0\rightarrow \pm \infty$, {or physically $|\epsilon_0| \gg \epsilon_\lambda,1/\epsilon_\lambda$}. Here we outline the main steps in this derivation. As is clear from Eqs. (\ref{xisymm}) and (\ref{chisymm}) for the symmetric case, the key quantities are the dimensionless functions $\mathcal F_0, \mathcal F_1, \mathcal G_0$ and $\mathcal G_1$

Starting with the limit $\epsilon_0\rightarrow \infty$, in the integrand for $\mathcal{F}_n$ we can approximate the term $\cosh^{-2}([\epsilon_0-x\epsilon_\lambda]/2) \approx 4e^{-(\epsilon_0-x\epsilon_\lambda)}$. For $\mathcal G_n$ we similarly approximate $\cosh^{-2}([\epsilon_0+y\epsilon_\lambda]/2)$ with $4e^{-(\epsilon_0+y\epsilon_\lambda)}$. We then have
\begin{eqnarray}
\mathcal{F}_0&\approx & 4e^{-\epsilon_0} \int_0^1 dx  ~\text{arcosh}\left(\frac{1}{\sqrt{x}}\right)e^{x\epsilon_\lambda}= e^{-\epsilon_0}\alpha(\epsilon_\lambda), \nonumber \\
\mathcal{F}_1&\approx& 4e^{-\epsilon_0} \int_0^1 dx  ~x~\text{arcosh}\left(\frac{1}{\sqrt{x}}\right)e^{x\epsilon_\lambda}=e^{-\epsilon_0}\frac{d\alpha(\epsilon_\lambda)}{d\epsilon_\lambda}, \nonumber \\
\mathcal{G}_0 &\approx & 4e^{-\epsilon_0} \int_0^\infty dy  ~\text{arsinh}\left(\frac{1}{\sqrt{y}}\right)e^{-y\epsilon_\lambda} = e^{-\epsilon_0}\beta(\epsilon_\lambda) \nonumber \\
\mathcal{G}_1 &\approx& 4e^{-\epsilon_0} \int_0^\infty dy  ~\text{arsinh}\left(\frac{1}{\sqrt{y}}\right)e^{-y\epsilon_\lambda} =e^{-\epsilon_0}\frac{d\beta(\epsilon_\lambda)}{d\epsilon_\lambda} \nonumber \\
\label{approxplusinf}
\end{eqnarray}
that is $\mathcal F_n, \mathcal G_n \sim e^{-\epsilon_0} \ll 1$, and we introduced
\begin{eqnarray}
\alpha(\epsilon_\lambda)&=& 4\int_0^1 dx  ~\text{arcosh}\left(\frac{1}{\sqrt{x}}\right)e^{x\epsilon_\lambda}, \nonumber \\ \beta(\epsilon_\lambda)&=& 4\int_0^\infty dy  ~\text{arcsinh}\left(\frac{1}{\sqrt{y}}\right)e^{-y\epsilon_\lambda},
\end{eqnarray}
Inserting the expressions in Eq. (\ref{approxplusinf}) into Eqs. (\ref{xisymm}) and (\ref{chisymm}) and expanding to leading order in $e^{-\epsilon_0}$ we get
\begin{equation}
\xi=\xi_\text{cl} + \xi^{+}_\text{qm}, \hspace{1cm} \chi= \chi^{+}_\text{qm},
\end{equation}
with the classical, geometrical term $\xi_\text{cl}=1/(1+c_g)$  and the quantum corrections
\begin{eqnarray}
 \xi^{+}_\text{qm}&=&e^{-\epsilon_0}h(\epsilon_\lambda)\frac{c_dc_g}{(1+c_g)^2},  \nonumber \\
 \chi^{+}_\text{qm}&=&\epsilon_0e^{-\epsilon_0}h(\epsilon_\lambda)\frac{c_d}{1+c_g},  
\end{eqnarray}
where for shortness we introduced $h(\epsilon_\lambda)=\alpha(\epsilon_\lambda)+\beta(\epsilon_\lambda)$. The quantum parts of the screening coefficients are thus exponentially suppressed for $\epsilon_0 \rightarrow \infty$.

In the opposite limit, $\epsilon_0\rightarrow -\infty$, in the integrand for $\mathcal{F}_n$ we can approximate $\cosh^{-2}([\epsilon_0-x\epsilon_\lambda]/2) \approx 4e^{(\epsilon_0-x\epsilon_\lambda)}$. This gives
\begin{eqnarray}
\mathcal{F}_0&\approx & 4e^{\epsilon_0} \int_0^1 dx  ~\text{arcosh}\left(\frac{1}{\sqrt{x}}\right)e^{-x\epsilon_\lambda}= e^{\epsilon_0}\alpha(-\epsilon_\lambda), \nonumber \\
\mathcal{F}_1&\approx& 4e^{\epsilon_0} \int_0^1 dx  ~x~\text{arcosh}\left(\frac{1}{\sqrt{x}}\right)e^{-x\epsilon_\lambda}=-e^{\epsilon_0}\frac{d\alpha(-\epsilon_\lambda)}{d\epsilon_\lambda}, \nonumber \\
\end{eqnarray}
For $\mathcal G_n$ we need to proceed slightly differently. Making use of the limiting expression $\mbox{arcsinh}(1/\sqrt{y})\approx 1/\sqrt{y}$ for large $y$ we first note that one can write the terms containing $\mathcal G_n$ in the numerator of $\chi$ in Eq.~(\ref{chisymm}) as
\begin{eqnarray}
&&\epsilon_0\mathcal{G}_0+\epsilon_\lambda \mathcal{G}_1 \nonumber \\
&=& \int_1^{\infty} dy ~(\epsilon_0+\epsilon_\lambda y)~\text{arsinh}\left(\frac{1}{\sqrt{y}}\right)\frac{1}{\cosh^2([\epsilon_0+y\epsilon_\lambda]/2)} \nonumber \\
&\approx & \epsilon_\lambda \int_{-\infty}^{\infty} dy' \frac{y'}{\sqrt{y'-\epsilon_0/\epsilon_\lambda}}\frac{1}{\cosh^2(\epsilon_\lambda y'/2)}  \nonumber \\
&\approx &  \epsilon_\lambda \int_{-\infty}^{\infty} dy' y'\left(\sqrt{\frac{\epsilon_\lambda}{-\epsilon_0}} -\frac{y'}{2}\left[ \frac{\epsilon_\lambda}{-\epsilon_0} \right]^{3/2}  \right) \frac{1}{\cosh^2(\epsilon_\lambda y'/2)}  \nonumber \\
&=&  -\epsilon_\lambda \frac{2\pi^2}{3\epsilon_\lambda^3}\left[ \frac{\epsilon_\lambda}{-\epsilon_0} \right]^{3/2}=-\frac{2\pi^2}{3\sqrt{\epsilon_\lambda}}\frac{1}{(-\epsilon_0)^{3/2}},
\end{eqnarray} 
expanding to leading order in $1/\epsilon_0$. Along the same lines we get
\begin{eqnarray}
\mathcal{G}_0=2\sqrt{\frac{\epsilon_\lambda}{-\epsilon_0}}\,.
\end{eqnarray} 
Inserting these results for $\mathcal F_0, \mathcal F_1, \mathcal G_0$ and $\mathcal G_1$  into Eqs. (\ref{xisymm}) and (\ref{chisymm}) and expanding again to leading order in $1/\epsilon_0$ (hence, the terms $e^{-\epsilon_0}$ do not contribute) we get
\begin{equation}
\xi=\xi_\text{cl} + \xi^{-}_\text{qm}, \hspace{1cm} \chi= \chi^{-}_\text{qm},
\end{equation}
with the quantum corrections
\begin{eqnarray}
  \xi^{-}_\text{qm}&=&2\sqrt{\frac{\epsilon_\lambda}{-\epsilon_0}}\frac{c_dc_g}{(1+c_g)^2}\,,   \nonumber \\ 
 \chi^{-}_\text{qm}&=& -\frac{1}{(-\epsilon_0)^{3/2}}\frac{2\pi^2}{3\sqrt{\epsilon_\lambda}} \frac{c_d}{1+c_g}\,.
\end{eqnarray}
%

%